\documentclass{jfm}
\usepackage[utf8]{inputenc}
\usepackage{graphicx}
\usepackage{float}
\usepackage{subcaption}
\usepackage{textcomp,gensymb} 
\usepackage{xcolor} 
\usepackage{amsmath} 
\usepackage{amssymb} 
\usepackage[colorlinks=true,citecolor=blue,linkcolor=blue]{hyperref}
\usepackage{natbib}

\newcommand{\mean}[1]{\left<#1\right>}
\newcommand{\ReNumber}{{Re}}

\newcommand{\MaNumber}{{Ma}}
\newcommand{\PeNumber}{{Pe}}
\newcommand{\BiNumber}{{Bi}}
\newcommand{\DaNumber}{{Da}}
\newcommand{\dif}{\mathrm{d}}
\newcommand{\itVector}[1]{\boldsymbol{#1}}
\newcommand{\boldcdot}{\boldsymbol{\cdot}}

\shorttitle{Slip of liquid-infused surfaces in the presence of surfactants}
\shortauthor{J. Sundin and S. Bagheri}

\title{Slip of submerged two-dimensional liquid-infused surfaces in the presence of surfactants}

\author{Johan Sundin\aff{1}\corresp{\email{johasu@mech.kth.se}} \and Shervin Bagheri\aff{1}}

\affiliation{\aff{1}FLOW Centre, Dept.~Engineering Mechanics, KTH, Stockholm SE-100 44, Sweden}

\begin{document}
\maketitle

\begin{abstract}
    Using numerical simulations, we investigate the effects of Marangoni stresses induced by surfactants on the effective slip length of liquid-infused surfaces (LIS) with transverse grooves. The surfactants are assumed soluble in the external liquid shearing the surface and can adsorb onto the interfaces. Two different adsorption models are used: a classical Frumkin model and a more advanced model that better describes the decrease of surface tension for minuscule concentrations. The simulations show that LIS may face even more severe effects of surfactants than previously investigated superhydrophobic surfaces. Constructing an analytical model for the effective slip length, we can predict the critical surfactant concentration for which the slip length decreases significantly. This analytical model describes both adsorption models of LIS on a unified framework if properly adjusted. We also advance the understanding of when surfactant advection gives rise to highly skewed interfacial concentrations -- the so-called partial stagnant cap regime. To a good approximation, this regime can only exist below a specific surfactant concentration given by the Marangoni number and the strength of the surfactants.
\end{abstract}

\section{Introduction}
Liquid-infused surfaces (LIS) are promising candidates for reducing drag, resisting biofouling and increasing heat transfer in liquid flows \citep{epstein12, solomon14, rosenberg16, sundin21}. These surfaces consist of a solid surface texture with a lubricating liquid that is immiscible with the external fluid. The fluid-fluid interfaces and mobility of the lubricant give rise to a slipping effect of the external flow. LIS can self-repair and are not sensitive to hydrostatic pressure, thereby being more robust than superhydrophobic surfaces (SHS) if properly designed \citep{wong11, wexler15b, sundin21roughness}. 

The functionality of LIS has been mostly characterized assuming perfectly clean external liquid flows. However, both in applications and in laboratory setups \citep{jacobi15, peaudecerf17}, LIS are exposed to trace amounts of surfactants. These are substances that can adsorb onto interfaces and alter the surface (or interfacial) tension. 
Surfactants influence phenomena such as foaming, wetting, dispersion, and emulsification, appearing in a large variety of products, e.g. 
cleaning agents, 
paints, 
cosmetics, 
pharmaceuticals, 
and motor oils 
\citep{rosen12}. 
It has recently been acknowledged that traces of surfactants may induce Marangoni stresses that counteract the slip of SHS \citep{peaudecerf17}. As we will demonstrate in this paper, the presence of surfactants in the system also modifies the slip of LIS. In applications where the mobility of the infusing liquid is crucial, it is necessary to understand the influence of surfactants on the performance, particularly, when measured slip lengths deviate significantly from their expected values.

Surfactants adsorbed onto interfaces accumulate at stagnation points when subjected to flow, building up concentration gradients and corresponding Marangoni stresses. The Marangoni stress counter-acts the shear stress from the overlying flow, reducing the slip length. Using numerical simulations, a significant slip length reduction of SHS has been observed at bulk concentrations $c_0 \approx 10^{-3}$ mol/m$^3$ using properties of the surfactant sodium dodecyl sulfate (SDS) \citep{peaudecerf17}.
Most recent experimental studies of SHS, which propose that surfactant gradients lead to slip degradation, have not added surfactants artificially \citep{kim12, bolognesi14, peaudecerf17, song18, temprano21}. 
Instead, the fluid systems have likely been contaminated by the surrounding environment. Indeed, it is generally accepted that surfactants appear as ``hidden variables'' since their concentrations are unknown  \citep{manikantan20}. Analytical models that relate the slip length to surfactant concentration are therefore crucial to interpreting experimental measurements.   
 
 \citet{landel20} developed an analytical theory to predict the effective slip length of SHS from relevant non-dimensional numbers in a two-dimensional channel flow with surfactants. The model assumed low concentrations and uniform interfacial concentration gradients. 
A high shear rate can result in the upstream part of an interface having almost no surfactant gradients, rendering the model inaccurate. Inspired by the analysis of buoyantly rising bubbles \citep{palaparthi06}, \citet{landel20} found that the surface could enter this regime when interfacial surfactant advection overcomes interfacial diffusion and bulk exchange rates. However, they were unable to find a quantitative condition for the transition. The analytical model compared favourably with numerical simulations for low shear rates, containing four fitted parameters. 
\citet{baier21} developed an analytical model for slip length degradation by insoluble surfactants. They considered a flow driven by an imposed shear stress and could therefore use the analytical flow solution of \citet{philip72a}.

The effects of surfactants on LIS have not been thoroughly investigated, although its importance has been acknowledged. Certain interfacial observations of LIS -- that cannot be fully explained -- have been attributed to the presence of surfactants. One example is the absence of interface deformations in the vicinity of stagnation points \citep{jacobi15}. 
The influence of surfactants on LIS drag reduction was also highlighted as a future challenge in a recent review \citep{hardt22}.
%
There are indications that LIS are more sensitive to surfactants compared to SHS. Dodecane, hexane and other alkanes are promising infusing liquids for drag reduction applications \citep{buren17}. However, the interfaces of aqueous surfactant solutions and saturated hydrocarbons (e.g.~alkanes) generally face a more significant decrease in surface tension than the interfaces of corresponding water-air systems (assuming only minor or no solubility of the surfactant in the hydrocarbon) \citep{rosen12}. 
The relatively early publication by \citet{gillap68} reported this effect for sodium decyl sulfate and SDS and various water-alkane interfaces (e.g.~water-hexane with SDS). 

Measurements have also shown that the surface tension of water-alkane interfaces can experience an initial decrease of several mN/m at minuscule surfactant concentrations. \citet{fainerman19} illustrated this effect for water-hexane interfaces with dodecyl and tridecyl dimethyl phosphine oxide (C$_{12}$DMPO and C$_{13}$DMPO, lowest concentration $c_0 = 10^{-6}$ mol/m$^3$). 
This effect is also present for several other surfactants such as SDS and trimethyl ammonium bromides (C$_{n}$TAB). 
It is only recently that adsorption models for water-oil interfaces have been developed to account for such phenomena \citep{fainerman20}.

Based on the current knowledge about surfactants at water-oil interfaces, we have investigated the dependency of the slip length of LIS on surfactant concentration. The coupled system of equations for the flow and the surfactants has been solved numerically for transverse grooves in laminar shear flow. 
In contrast to SHS, it is necessary to resolve the flow both inside and outside the textures of LIS. We also developed an analytical model for the slip length in the presence of surfactants. The model can be used to predict the reduction of slip lengths if the surfactant type and concentration in bulk liquid are known. In settings where  surfactants are hidden in the system, the model may be used to estimate the concentration of surfactants given measurements of the slip length.

The flow configuration, governing equations, and numerical methods are described in the next section. Sec.~\ref{sec:analyticalViscousInfusedLiquid} introduces the analytical model for an imposed Marangoni stress. Interfacial surfactant adsorption and desorption of surfactants using regular Frumkin kinetics are described and used for simulations in sec.~\ref{sec:regularFrumkin}, followed by the corresponding analytical model (sec.~\ref{sec:analyticalSlipModel}). A more advanced model giving consistent surface tensions at minuscule concentrations is introduced in sec.~\ref{sec:advancedFrumkin}. Higher applied shear stresses, resulting in highly skewed interfacial surfactant concentrations (the partial stagnant cap regime), are treated in sec.~\ref{sec:stagnantCap}. Final remarks and conclusions are presented in secs.~\ref{sec:remarks} and \ref{sec:conclusions}, respectively.

\section{Configuration, governing equations, and numerical method}
\label{eq:governingEquations}
\begin{figure}
  \centering 
  \vspace{0.1cm}
  \includegraphics[height=5.5cm]{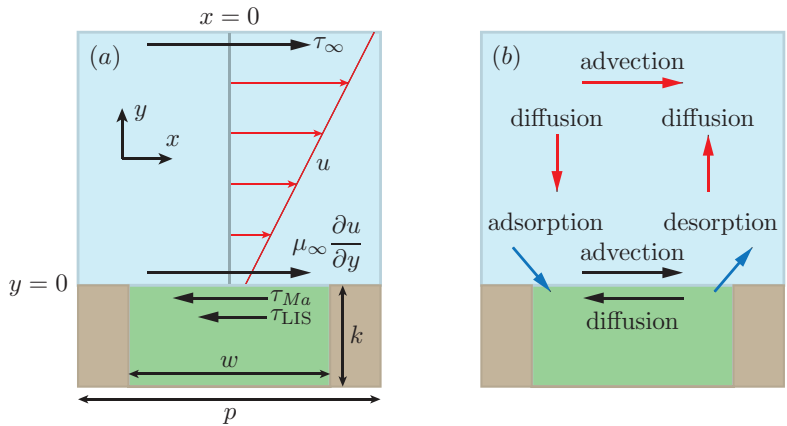} 
  \begin{subfigure}{\textwidth} 
    \captionlistentry{} \label{fig:shearBalance} 
  \end{subfigure}
  \begin{subfigure}{\textwidth} 
    \captionlistentry{} \label{fig:surfactantTransport} 
  \end{subfigure}
  \caption{ ($a$) Interface unit cell, illustrating the balance of stresses at the interface. The velocity profile of the external flow is also shown (red). ($b$) The physical processes transporting surfactants in the bulk and on the interface and controlling their exchange (red, black, and blue arrows, respectively). The external liquid is blue, the infusing liquid green, and the solid brown. }
  \label{fig:schematics}
\end{figure}

We consider a LIS texture consisting of a periodic array of rectangular transverse grooves subjected to steady laminar flow. Due to the texture periodicity, it is sufficient to consider one interface unit cell with a single groove, sketched in fig.~\ref{fig:schematics}a. The groove depth was $k = 50$ \textmu m, the width $w = 2k$, and the pitch $p = 3k$ (groove centre-to-centre distance). 

The external and infusing fluids have viscosities $\mu_\infty$ and $\mu_i$, respectively ($\mu_\infty = 1.0$ mPas for water), 
forming interfaces aligned with the ridges between the grooves.
Flat, non-deformable interfaces are assumed in this study so that the sole effect of the surfactants is the Marangoni force. An increased deformation due to a reduction in surface tension is expected to be of secondary importance \citep{landel20}.

The flow velocities and texture dimensions are small. Therefore, we use the Stokes equations for a steady incompressible flow,
\begin{equation}
    0 = -\nabla P +  \nabla \boldcdot \mu\left(\nabla \itVector{u} + (\nabla \itVector{u})^\mathrm{T}\right) 
    \quad \text{ and } \quad
    \nabla \boldcdot \itVector{u} = 0,
    \label{eq:momentumTransport}
\end{equation}
where $P$ is the pressure, $\mu$ is the fluid viscosity, and $\itVector{u} = (u,v)$ is the fluid velocity with streamwise and wall-normal components $u$ and $v$, respectively. 
The streamwise coordinate is $x$ with $x = 0$ in the centre of the considered fluid-fluid interface, and $y$ is the wall-normal coordinate with $y = 0$ at the interface.
Eqs.~\eqref{eq:momentumTransport} are valid for both the external and infusing fluids. No-slip and impermeability conditions ($\itVector{u} = 0$) were used at solid boundaries. At the interface, the wall-normal velocity was $v = 0$ and the streamwise velocity $u$ was assumed to be continuous. 
The balance of shear stress on the interface is \citep{leal07} 
\begin{equation}
    \mu_\infty\left.\frac{\partial u}{\partial y}\right|_{y = 0^+} = \mu_i\left.\frac{\partial u}{\partial y}\right|_{y = 0^-} - \frac{\dif \gamma}{\dif x}, 
    \label{eq:shearBC}
\end{equation}
where $\gamma$ is the surface tension. The velocity gradients in eq.~\eqref{eq:shearBC} have been evaluated precisely above and below the interface ($y = 0^+$ and $0^-$, respectively). The imposed shear stress that drives the flow is assumed to be $\tau_\infty$ at $y \to \infty$.

In order to simplify subsequent notation, we introduce 
\begin{equation}
    \tau_\mathrm{LIS} = \mu_i\left.\frac{\partial u}{\partial y}\right|_{y = 0^-} \quad \text{ and } \quad \tau_\MaNumber = -\frac{\dif \gamma}{\dif x}.
\end{equation}    
The balance of stresses is shown in fig.~\ref{fig:shearBalance}. 
For no surfactants ($\tau_\MaNumber = 0$), eq.~\eqref{eq:shearBC} relaxes to the classical interface condition, and for $\tau_\mathrm{LIS} = 0$, ideal (gas infused) SHS are regained. 

\begin{figure}
  \centering 
  \includegraphics[height=3.5cm]{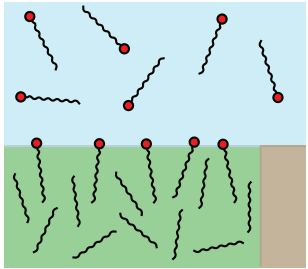} 
  \caption{ Schematic illustration of surfactants dissolved in water and adsorbed at a water-alkane interface. The interface is shown pinned to a corner of the solid texture. SDS and C$_{12}$TAB molecules consist of hydrocarbon tails with 12 carbon atoms and hydrophilic head groups (red). Alkane molecules are also hydrocarbon chains. Dodecane and hexane molecules have $12$ and $6$ carbon atoms, respectively, here assumed to be arranged in straight chains. The different phases have the same colours as in fig.~\ref{fig:schematics}. }
  \label{fig:alkaneSurfactantInteraction}
\end{figure}

The surfactants are assumed to be soluble in the water phase and can adsorb at the interfaces, (see sketch in fig.~\ref{fig:alkaneSurfactantInteraction}). Surfactant transport mechanisms are illustrated in fig.~\ref{fig:surfactantTransport}. An advection-diffusion equation governs the interfacial surfactant concentration $\Gamma$,
\begin{equation}
    \frac{\dif}{\dif x}(u_s \Gamma) = D_s\frac{\dif^2 \Gamma}{\dif x^2} + S,
    \label{eq:interfacialTransport}
\end{equation}
where $u_s$ is the velocity of the interface plane ($y = 0$), $D_s$ is the interface diffusivity, and $S$ is a source term describing adsorption and desorption. The source term would be zero for insoluble surfactants. No-flux conditions apply at interface edges ($\dif \Gamma/\dif x = 0$), and eq.~\eqref{eq:interfacialTransport} is valid for $-w/2 < x < w/2$. The surface coverage of the interfacial surfactants is $\theta = \Gamma/\Gamma_m = \omega \Gamma$, where $\Gamma_m$ is the maximum possible interface concentration and $\omega$ is the molar area. To increase brevity of the expressions throughout the paper, we will use non-dimensional surfactant concentration, $\theta$, but the theory is otherwise presented using dimensional quantities.

The equation governing bulk surfactant concentration $c$ is 
\begin{equation}
     \nabla \boldcdot (\itVector{u}c) = D\nabla^2 c,
     \label{eq:bulkTransport}
\end{equation}
where $D$ is the (bulk) diffusivity. The adsorption and desorption of surfactants at an interface balance the diffusive flux:
\begin{equation}
    D\left.\frac{\partial c}{\partial y}\right|_{y = 0} = S. 
    \label{eq:bulkTransportBC}
\end{equation}
At solid boundaries, $\partial c/\partial y = 0$. Sufficiently far above the interface, we assume a constant surfactant concentration $c_0$. The diffusivities were set to $D = D_s = 7.0 \cdot 10^{-10}$ m$^2$/s, which also was used by \citet{peaudecerf17}.

The source term and the Marangoni stresses are described in the following sections. We adopt surface tension and adsorption/desorption models for the two extensively studied surfactants SDS and C$_{12}$TAB 
on water-air \citep{peaudecerf17} and water-alkane interfaces \citep{fainerman19}. 
SDS can be found in personal care products, 
and C$_{12}$TAB can, for example, be used to stabilise foam \citep{rosen12, carey09}.
These surfactants are illustrated schematically in fig.~\ref{fig:alkaneSurfactantInteraction}. They have hydrophobic hydrocarbon tails with 12 carbon atoms but different hydrophilic (head) groups.

\subsection{Numerical method}
The numerical simulations were performed using the finite element solver FreeFem++ \citep{hect12, lacis20}. Velocities and surfactant concentrations were discretised using quadratic ($P_2$) finite elements while linear ($P_1$) elements were used for the pressure. The system of equations (\ref{eq:momentumTransport}, \ref{eq:interfacialTransport}, \ref{eq:bulkTransport}) was solved iteratively. The domain consisted of one interface unit cell (fig.~\ref{fig:schematics}). The external flow domain had a height of $3k$. At the upper boundary, we imposed constant shear stress $\tau_\infty$, zero wall-normal stress, and bulk surfactant concentration $c_0$. Periodic boundary conditions were applied in the streamwise direction. 

The computational mesh used for the flow (bulk and cavity) and the bulk surfactants ($y \ge 0$) was generated using a built-in tool. The cells were triangular, and the number of cells was prescribed at the different domain boundaries with a spacing of $w/N$, where $N = 64$. The grid was refined close to the interface. For the simulations with a low flow speed (secs.~\ref{sec:regularFrumkin} and \ref{sec:advancedFrumkin}), two refinements were made, reducing the sides of the cells by a factor of four in total, i.e. $N = 256$. The simulations with artificially applied Marangoni stress (sec.~\ref{sec:analyticalViscousInfusedLiquid}) also had $N = 256$ at the interface. For simulations with high flow speed, we used four refinements ($N = 1024$, sec.~\ref{sec:stagnantCap}). We have also performed grid refinement studies, see appendix \ref{sec:gridStudy}.

The interfacial surfactant transport equation \eqref{eq:interfacialTransport} was solved on a one cell high mesh with equal cell spacing to the flow and bulk surfactant mesh in the streamwise direction. We imposed periodic boundary conditions in the $y$-direction. The interface velocity and the bulk surfactant concentration appearing in the source term ($u_s$ and $c_s$, respectively) were taken at $y = 0$. There were no variations in the $y$-direction in the solution of the interfacial surfactant concentration; it exactly corresponds to the one-dimensional solution.

\section{Analytical model of a viscous infusing liquid}
\label{sec:analyticalViscousInfusedLiquid}
To construct an analytical model of the flow over LIS with surfactants, we assume that the stresses $\tau_\mathrm{LIS}$ and $\tau_\MaNumber$ are constant. Essentially, we combine the models presented by \citet{schonecker14} and \citet{landel20}. The former considered the flow over LIS without surfactants and assumed  $\tau_\mathrm{LIS}$ was constant. The latter work considered flow over SHS and modelled the Marangoni stresses $\tau_\MaNumber$ as constant. In this section, we present the analytical model of our system and refer the reader to appendix \ref{sec:analyticalExpressionVelTransGroove} for derivations.

The (effective) slip length $b$ is defined by
\begin{equation}
    U_s = b\left.\mean{\frac{\partial u}{\partial y}}\right|_{y = 0^+} = b\frac{\tau_\infty}{\mu_\infty},
\end{equation}
where $\mean{}$ gives the average in the streamwise direction ($-p/2 < x \le p/2$), and $U_s = \mean{u}$ at $y = 0$ is the slip velocity.
The resulting analytical expression of the effective slip length is (eq.~\ref{eq:derivedSlipRelation})
\begin{equation}
    b =
    b_\mathrm{SHS}\beta_\mathrm{LIS} \left(1 - \frac{\tau_\MaNumber}{\tau_\infty}\right),
    \label{eq:slipRelation}
\end{equation}
where we have introduced the ideal SHS slip length ($\mu_i/\mu_\infty = \tau_\MaNumber/\tau_\infty = 0$)
\begin{equation}
    b_\mathrm{SHS} = -\frac{p}{2\pi}\ln(\cos \alpha),
    \label{eq:idealSHSSlipLengthRelation}
\end{equation}
with $\alpha = (\pi/2)(w/p)$. The factor $\beta_\mathrm{LIS} = C_t/(1 + C_t)$ describes the effects due to the viscous infusing liquid, where $C_t$ depends on the groove geometry and the viscosity ratio,
\begin{equation}
    C_t = \frac{8\alpha D_t\mu_\infty/\mu_i}{\ln\left(\dfrac{1 + \sin(\alpha)}{1 - \sin(\alpha)}\right)}.
\end{equation}
The parameter $D_t$ is a normalised maximum local slip length (eq.~\ref{eq:DtModel}). Eq.~\eqref{eq:slipRelation} describes the slip length as a linear function of the Marangoni stress. Similarly, the velocity at the centre of the interface is (eq.~\ref{eq:derivedMiddleInterfaceVel})
\begin{equation}
    u_{s}^0 = u_{s,\mathrm{SHS}}^0\beta_\mathrm{LIS}\left(1 - \frac{\tau_\MaNumber}{\tau_\infty}\right),
    \label{eq:middleInterfaceVel}
\end{equation}
where 
\begin{equation}
   u_{s,\mathrm{SHS}}^0 = \frac{\tau_\infty w}{4\mu_\infty\alpha}\mathrm{arccosh}\left( \frac{1}{\cos\alpha}\right).
   \label{eq:uSRelation}
\end{equation} 

\begin{figure}
  \centering
  \includegraphics[width=5.5cm]{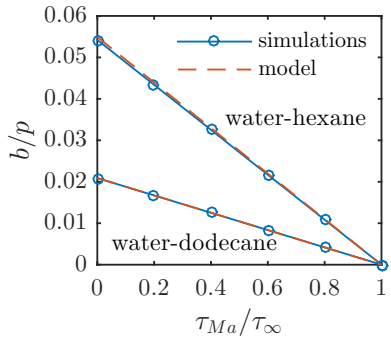} 
  \caption{Comparison of eq.~\eqref{eq:slipRelation} to simulation results where $\tau_\MaNumber$ was applied  artificially. The viscosity ratio was set to $\mu_i/\mu_\infty = 1.4$ or $0.33$ (corresponding to water-dodecane or water-hexane, respectively). }
  \label{fig:slipLengthModelValidation}
\end{figure}

What remains is to find an explicit expression for $\tau_\MaNumber$. However, we can compare the expression for $b$ in \eqref{eq:slipRelation} to simulations of eqs.~\eqref{eq:momentumTransport} and \eqref{eq:shearBC} with an artificially applied $\tau_\MaNumber$. A comparison is shown in fig.~\ref{fig:slipLengthModelValidation}, with a convincing agreement. The two examined viscosity ratios were $\mu_i/\mu_\infty = 1.4$ and $0.33$, corresponding to water-dodecane and water-hexane interfaces, respectively (tab.~\ref{tab:surfactantProperties}). For both viscosity ratios, $b$ decreases from $b_\mathrm{SHS}\beta_\mathrm{LIS}$ to $0$ when $\tau_\MaNumber/\tau_\infty$ increases from $0$ to $1$, as predicted by eq.~\eqref{eq:slipRelation}. The resulting slip length is higher for water-hexane interfaces (for the same $\tau_\MaNumber/\tau_\infty$) because of the lower viscosity ratio.

\section{Adsorption and desorption with regular Frumkin kinetics}
\label{sec:regularFrumkin}
The interfacial adsorption and desorption rates determine the source term $S$. 
These rates have in recent studies of SHS been modelled by Frumkin kinetics \citep{peaudecerf17, landel20}, 
which are consistent with the Frumkin isotherm \citep{chang95}. 
Therefore, we also adopt Frumkin kinetics for this investigation. The source term becomes
\begin{align}
    S &= \frac{1}{\omega}\left(\kappa_{a} c_s\left(1 - \theta\right) - \kappa_{d}\theta e^{-2a\theta}\right),
    \label{eq:simpleSource}
\end{align}
where $c_s$ is the value of $c$ at the interfaces (dependent on $x$), $a$ is a constant, $\kappa_{a}$ is the adsorption coefficient, and $\kappa_{d}$ is the desorption coefficient. If the surfactant molecules are mutually attractive, $a$ is positive, and if they are repulsive, $a$ is negative \citep{manikantan20}. 
In equilibrium, adsorption and desorption fluxes balance (i.e. $S = 0$), and it is sufficient to state $\kappa_{a}/\kappa_{d}$ instead of the individual values of $\kappa_{a}$ and $\kappa_{d}$. 

The expression for the surface tension (equation of state) corresponding to the Frumkin isotherm is
\begin{equation}
    \gamma = \gamma_c + \frac{nRT}{\omega}\left[\ln(1 - \theta) + a{\theta}^2\right],
    \label{eq:surfaceTension}
\end{equation}
where $\gamma_c$ is the surface tension for $\theta = 0$, $R = 8.314~\mathrm{J}/(\mathrm{K} \mathrm{mol})$ is the universal gas constant, $T = 293~\mathrm{K}$ (at 20\degree C) is the absolute temperature, and $n$ is a constant. The factor $n$ can account for the adsorption of counter-ions \citep{chang95, fainerman19}. Without a supporting electrolyte, SDS has $n = 2$, which is the value used here. We assume that the data used for C$_{12}$TAB was obtained using a solution with a sufficiently high concentration of supporting electrolyte so that $n = 1$. 
It is generally accepted that the equilibrium model eq.~\eqref{eq:surfaceTension} can be used in  non-equilibrium conditions \citep{chang95}. 

\begin{table}
    \centering
    \begin{tabular}{lllll}
    Surfactant                                         & SDS               & SDS               & SDS               & C$_{12}$TAB  \vspace{0.2cm} \\ 
    Interface                                          & water-air         & water-dodecane    & water-dodecane    & water-hexane \\
    $\mu_i/\mu_\infty$                                 & 0.01              & 1.4               & 1.4               & 0.33         \\
    Ads./des.~model                                    & reg.~Frumkin      & reg.~Frumkin      & adv.~Frumkin      & adv.~Frumkin \\  
    $n$                                                & 2                 & 2                 & 2                 & 1            \\
    $\omega$   [$10^5$ m$^2$/mol]                      & 2.551             & 5.6               & 4.0               & 4.1          \\
    $a$                                                & 1.2               & 0.9               & 0.9               & 0            \\ 
    $\kappa_a/\kappa_d$ [m$^3$/mol]                    & $0.179$           & $2.7$             & $1.8$             & $59$         \\ 
    $\kappa_d$ [1/s]                                   & 500               & 500 and 5         & 5                 & 5            \\ 
    $\omega^a$ [$10^5$ m$^2$/mol]                      & --                & --                & 3.5               & 3.5          \\
    $a^{as}$                                           & --                & --                & 0.9               & 1.0          \\ 
    $a^a$                                              & --                & --                & 0                 & 0            \\
    $c_0^a$ [$10^3$ mol/m$^3$]                         & --                & --                & 4.4               & 6.8          \\
    $\kappa_{a,0}^a/\kappa_d^a$ [m$^3$/mol]            & --                & --                & $1.5\cdot10^{-2}$ & $6.5\cdot10^{-3}$ \\
    $\kappa_{a,\mathrm{max}}^a/\kappa_d^a$ [m$^3$/mol] & --                & --                & $3.5\cdot10^{-4}$ & $8.0\cdot10^{-5}$   \\
    \end{tabular}
    \caption{Parameters used for SDS at water-air and water-dodecane interfaces and C$_{12}$TAB at water-hexane interfaces. Reg.~Frumkin refers to the model expressed by eqs.~\eqref{eq:simpleSource} and \eqref{eq:surfaceTension} and adv.~Frumkin to eqs.~\eqref{eq:advSource}, \eqref{eq:alkaneSource} and \eqref{eq:advSurfaceTension}. For all setups, we assume $D = D_s = 7.0 \cdot 10^{-10}$ m$^2$/s. } 
    \label{tab:surfactantProperties}
\end{table}

\begin{figure}
  \centering
      \includegraphics[height=4.5cm]{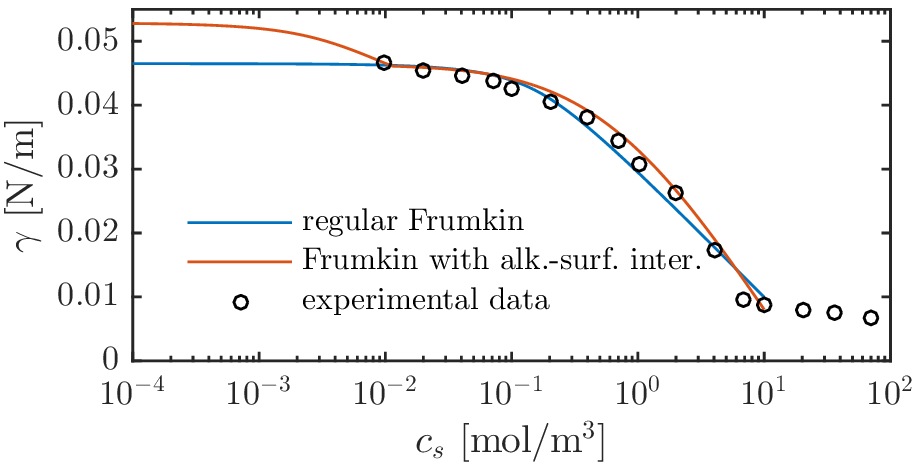} 
  \caption{Equilibrium surface tension ($S = 0$) of SDS on a water-dodecane interface, comparing experimental data of \cite{fainerman19} to the regular Frumkin model and a Frumkin model taking into account alkane-surfactant interaction. The surface tension is $\gamma_c = 52.87$ mN/m for water-dodecane without surfactants at 20\degree C \citep{zeppieri01}. In contrast to water-air interfaces, the regular Frumkin model does not exhibit the correct asymptotic behaviour when $c_s \to 0$ at water-oil interfaces. }
  \label{fig:equiSurfTenSDSExp}
\end{figure}

We performed simulations of SDS on water-dodecane interfaces with regular Frumkin kinetics (eqs.~\ref{eq:simpleSource} and \ref{eq:surfaceTension}). We increased the concentration $c_0$ in steps starting from $10^{-8}$ mol/m$^3$. Equilibrium parameters were adapted from \citet{fainerman19}, compensated for not explicitly considering interactions of surfactant-alkane molecules (sec.~\ref{sec:advancedFrumkin}). These parameters are summarised in tab.~\ref{tab:surfactantProperties} and result in an equilibrium-state surface tension shown in fig.~\ref{fig:equiSurfTenSDSExp}. The figure also contains experimental data from \citet{fainerman19}. Regular Frumkin kinetics can describe the measurements but not the correct asymptotic behaviour at minuscule concentrations. 

Next, we consider the non-equilibrium system by imposing an external shear stress. 
As a reference to SDS on water-dodecane interfaces, we also simulated SDS on water-air interfaces, starting at $c_0 = 10^{-7}$ mol/m$^3$. Air-water parameters for equilibrium were taken from \citet{prosser01}. 
The relatively low imposed shear stress was $\tau_\infty = 0.33$ mPa; the effects of increased shear stress are described in sec.~\ref{sec:stagnantCap}. 
The non-equilibrium parameter $\kappa_d = 500$ 1/s of the air-water interfaces corresponds to the value reported by \cite{chang95}. This value was determined using empirically modified Langmuir-Hinshelwood kinetics. This model allows $\kappa_d$ to vary with concentration and includes an exponential factor in the adsorption flux similar to the desorption flux (cf.~\ref{eq:simpleSource}). The value used here corresponds to the lowest bulk concentration (1.7 mol/m$^3$). 
For the water-dodecane system, we used both $\kappa_d = 500$ and $5$ 1/s, with minor differences. The latter implies a lower desorption rate. For all other simulations, we used $\kappa_d = 5$ 1/s. 

\begin{figure}
  \centering 
  \begin{subfigure}{0.48\textwidth}
      \centering
      \includegraphics[height=5.5cm]{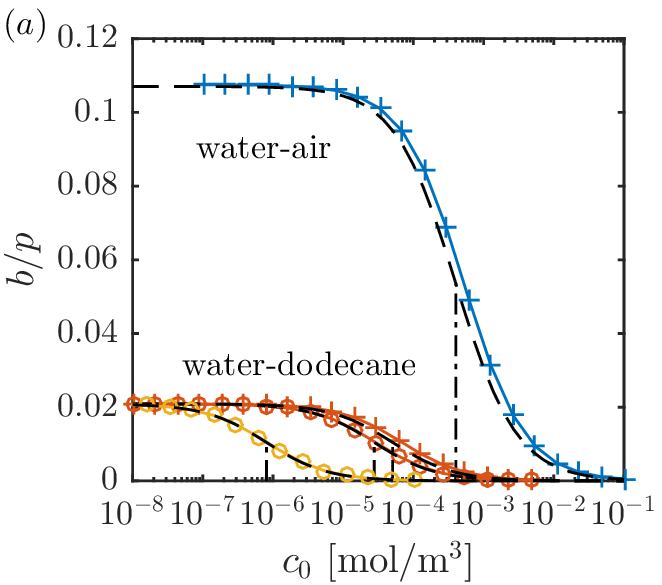} 
       \captionlistentry{}
      \label{fig:frumkinSDSComparison}
  \end{subfigure}
  \begin{subfigure}{0.48\textwidth}
      \centering
      \includegraphics[height=5.5cm]{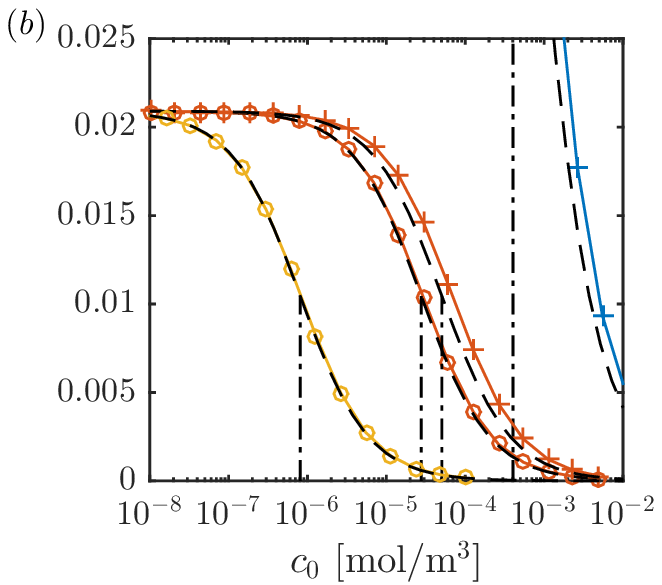} 
      \captionlistentry{}
      \label{fig:frumkinSDSComparisonZoomIn}
  \end{subfigure}
  \caption{ Parametric study of slip length at $\tau_\infty = 0.33$ mPa for different bulk concentrations of SDS with a water-air (blue) and a water-dodecane interface (red regular model, yellow advanced model), together with the analytical model (dashed lines). SHS and LIS results are shown in ($a$), and a zoom-in of the LIS results in ($b$). Symbols refer to $\kappa_d = 5$ (\opencirc) and $500$ ($+$) 1/s. The vertical dashed-dotted lines mark $\alpha_\mathrm{diff} + \alpha_S = 1$, corresponding to $b/(b_\mathrm{SHS}\beta_\mathrm{LIS}) \approx 1/2$ (eq.~\ref{eq:slipLengthEstimation}).}
  \label{fig:frumkinSDSResults}
\end{figure}

Fig.~\ref{fig:frumkinSDSComparison} shows the resulting slip lengths as a function of $c_0$, and fig.~\ref{fig:frumkinSDSComparisonZoomIn} provides a zoomed-in view. The water-air and water-dodecane results are shown in blue and red, respectively. 
Results for the advanced adsorption model are also included in the figure (yellow markers, see sec.~\ref{sec:advancedFrumkin}). The slip length is about five times higher for the SHS without surfactants ($c_0 = 0$ mol/m$^3$). As $c_0$ increases, the slip length of both systems decreases. The bulk concentration giving a significantly decreased slip length ($b/(b_\mathrm{SHS}\beta_\mathrm{LIS}) \approx 0.5$) is around one order of magnitude smaller for the water-dodecane than the water-air system, marked by vertical dashed-dotted lines in fig.~\ref{fig:frumkinSDSResults}. For the water-air system and the water-dodecane systems ($\kappa_d = 500$ 1/s), these concentrations are $c_0 = 4\cdot10^{-4}$ and $5\cdot10^{-5}$ mol/m$^3$, respectively. This difference reflects the equilibrium surface tension behaviour. In order to explain the results in more detail, we continue to develop the analytical model of sec.~\ref{sec:analyticalViscousInfusedLiquid} in sec.~\ref{sec:analyticalSlipModel}. This analytical model is also included in fig.~\ref{fig:frumkinSDSResults} (dashed lines).

The analytical model predicts that the slip length becomes independent of $\kappa_d$ when this parameter is sufficiently large. At these desorption rates, diffusion of bulk surfactants becomes the limiting process (Damk\"ohler number $\DaNumber_\delta \gg 1$). \citet{temprano21} recently discussed this independence, using $\kappa_d = 0.75$ 1/s with good agreement to experimental results of SHS with surfactants naturally occurring in their experimental setting.

\section{Analytical model of surfactant transport}
\label{sec:analyticalSlipModel}

\begin{figure}
  \centering
      \includegraphics[height=5.0cm]{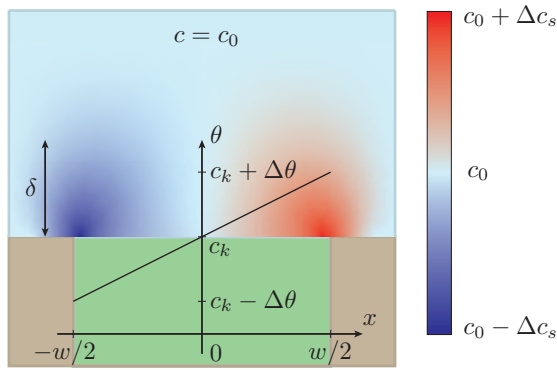} 
  \caption{Illustration of the distributions of bulk and interfacial surfactant concentrations ($c$ and $\theta$, respectively) of the analytical model. It is assumed that $c$ and $\theta$ vary linearly over the interfaces. The infusing liquid and the solid colours are the same as in fig.~\ref{fig:schematics}. }
  \label{fig:analyticalModelAssumptions}
\end{figure}

The analytical model of surfactant transport discussed here uses similar core assumptions as previous works \citep{landel20}. It is assumed that interfacial concentrations are low ($\theta \ll 1$) so that the governing equations \eqref{eq:simpleSource} and \eqref{eq:surfaceTension} can be linearised. A second assumption is that bulk and interfacial surfactant concentrations vary linearly over the interfaces (fig.~\ref{fig:analyticalModelAssumptions}). Such distributions imply approximately constant surface tension gradients, i.e.~the uniformly retarded regime. 

\subsection{Modelling bulk exchange}
\label{sec:modellingBulkExchange}
For a steady flow, there is a balance between adsorption and desorption processes. The integral of eq.~\eqref{eq:bulkTransportBC} over the interfaces must then be zero,
\begin{equation}
    D\int_{-w/2}^{w/2}\left.\frac{\partial c}{\partial y}\right|_{y = 0} \dif x = \int_{-w/2}^{w/2}S \dif x = 0.
\end{equation}
This condition implies the existence of a point $x_0$ on an interface where $D\partial c/\partial y |_{y = 0} = S = 0$. Since the wall-normal derivative of $c$ is zero at this location, we have $c_s \approx c_0$. 
The linearised source term (eq.~\ref{eq:simpleSource}) is
\begin{equation}
    S \approx \frac{\kappa_{d}}{\omega}\left(c_{k} \frac{c_s}{c_0} - \theta\right),
    \label{eq:simpleSourceNonDimlinearised}
\end{equation}
where we have introduced the non-dimensional bulk surfactant concentration $c_k = \kappa_ac_0/\kappa_d$. From the linearised source term,
\begin{equation}
    \theta \approx c_k \quad \text{ at } \quad x = x_0.
    \label{eq:linThetaValue}
\end{equation}
Due to advection, $\theta$ decreases upstream and increases downstream of $x_0$. It is assumed that $x_0 \approx 0$, that $\theta$ varies around $c_{k}$ by $\pm\Delta \theta$, and in the same way $c_s$ around $c_0$ by $\pm\Delta c_s$. These assumed concentration distributions imply (eqs.~\ref{eq:bulkTransportBC} and \ref{eq:simpleSourceNonDimlinearised}),
\begin{equation}
    \left.\frac{\partial c}{\partial y}\right|_{y = 0} \approx \frac{\kappa_d c_k}{D \omega}\left(\frac{\Delta \theta}{c_k} - \frac{\Delta c_s}{c_0}\right)\frac{-x}{w/2} = \frac{c_0}{w}\DaNumber\left(\frac{\Delta \theta}{c_k} - \frac{\Delta c_s}{c_0}\right)\frac{-x}{w/2},
    \label{eq:DeltacEstimation}
\end{equation}
where $\DaNumber = \kappa_a w/(\omega D)$ is the Damk\"ohler number \citep{temprano21}. 

From eq.~\eqref{eq:simpleSourceNonDimlinearised}, a characteristic adsorption flux is given by 
\begin{equation}
    j_a = \kappa_d c_k c_s/(\omega c_0) = \kappa_a c_s/\omega \sim \kappa_a c_0/\omega,
    \label{eq:bulkAdsorptionRate}
\end{equation}
in mol/(sm$^2$). Eq.~\eqref{eq:linThetaValue} entails that the characteristic desorption flux is the same: $\kappa_d \theta/\omega \sim \kappa_d c_k/\omega = \kappa_a c_0/\omega$. A corresponding scale for the diffusive flux of bulk surfactants is $D c_0/w$ (eq.~\ref{eq:bulkTransportBC}). Hence, $\DaNumber$ expresses characteristic adsorption/desorption flux to diffusive flux of bulk surfactants. Corresponding rates are found by multiplication by $\omega$. The actual values of $\Delta \theta$ and $\Delta c_s$ are neglected, but we do include the characteristic sizes of $\theta$ and $c_s$. This interpretation of $\DaNumber$ is also seen in eq.~\eqref{eq:DeltacEstimation}; large diffusion flux and low adsorption result in smaller wall-normal derivative of $c$. 

As pointed out by \citet{landel20}, eq.~\eqref{eq:DeltacEstimation} can be used to estimate the boundary layer thickness $\delta$ of $c$. Approximating $\delta$ by
\begin{equation}
    \left.\frac{\partial c}{\partial y}\right|_{y = 0, x = -w/2} \approx \frac{\Delta c_s}{\delta}  \implies \frac{\Delta c_s}{c_0} \approx \frac{\Delta \theta}{c_k}\frac{\DaNumber \frac{\delta}{w}}{1 + \DaNumber \frac{\delta}{w}}.
    \label{eq:cGradScaling}
\end{equation}
The modified Damk\"ohler number $\DaNumber_\delta = \DaNumber \delta/w$ considers the diffusion length scale of bulk surfactants to be $\delta$, which is more appropriate than $w$ \citep{palaparthi06}. We see that if the adsorption/desorption rate is much larger than the diffusion rate ($\DaNumber_\delta \gg 1$), $\Delta c_s/c_0 \approx \Delta \theta/c_k$. Otherwise, if $\DaNumber_\delta \ll 1$, $\Delta c_s/c_0 \approx \DaNumber_\delta \Delta \theta/c_k$.

The bulk surfactant transport equation \eqref{eq:bulkTransport} implies that $\delta$ depends on $w$ and P\'eclet number $\PeNumber = Uw/D$, where $U = w\tau_\infty/\mu_\infty$ is the characteristic velocity at $y = w$. Diffusion between boundary layers of adjacent grooves also introduces a dependency on $p$. An analytical estimate (appendix \ref{sec:derivationOfDelta}) resulted in 
\begin{equation}
    \frac{\delta}{w} = \frac{\frac{1}{2}\sqrt{1 - \frac{w}{p}}}{\left(1 + \frac{2}{3}\PeNumber\left(\frac{1}{2}\sqrt{1 - \frac{w}{p}}\right)^3\right)^{1/3}}.
    \label{eq:delta}
\end{equation}
The left relation of \eqref{eq:cGradScaling} was also used to compute $\delta$ explicitly. It was found that $\delta/w = 0.3$ was a relatively good approximation for all tested configurations for the current geometry and $\PeNumber$ (appendix \ref{sec:derivationOfDelta}, fig.~\ref{fig:deltaEstimation}a,b), in agreement with eq.~\eqref{eq:delta}. Deviations were around $\pm 0.1$.

\subsection{Interfacial surfactant transport balance}
\label{sec:surfactantTransportBalance}
The transport equation for the interfacial surfactant concentration \eqref{eq:interfacialTransport} expresses a balance between advection, diffusion and adsorption/desorption (compare fig.~\ref{fig:surfactantTransport}). We integrate this equation from $x = -w/2$ to $x = x_0$ and use the boundary condition \eqref{eq:bulkTransportBC},
\begin{equation}
    \left.\left( u_s\theta\right)\right|_{x=x_0} = D_s\left.\frac{\dif \theta}{\dif x}\right|_{x=x_0} + D\omega\int_{-w/2}^{x_0} \left.\frac{\partial c}{\partial y}\right|_{y = 0} \dif x.
    \label{eq:interfacialTransportIntegated}
\end{equation}
Using $x_0 \approx 0$ and eq.~\eqref{eq:cGradScaling} with the streamwise dependency \eqref{eq:DeltacEstimation}, we have
\begin{equation}
    \int_{-w/2}^{x_0} \left.\frac{\partial c}{\partial y}\right|_{y = 0} \dif x \approx  \frac{c_0}{w}\int_{-w/2}^{0} \frac{\Delta \theta}{c_k}\frac{\DaNumber}{1 + \DaNumber\frac{\delta}{w}}\frac{-x}{w/2} \dif x = c_0\frac{1}{4}\frac{\Delta \theta}{c_k}\frac{\DaNumber}{1 + \DaNumber_\delta}.
\end{equation}
With eq.~\eqref{eq:linThetaValue} and $\left.\dif \theta/\dif x\right|_{x=x_0} \approx 2\Delta \theta /w$, we get an expression for the interface centre velocity, 
\begin{equation}
   u_s^0 \approx U\left(c_1 \frac{1}{\PeNumber_s} + c_2\frac{\BiNumber}{1 + \DaNumber_\delta}\right)\frac{\Delta \theta}{c_k}
   \label{eq:interfaceVel2}
\end{equation}
with coefficients $c_1 \approx 2$ and $c_2 \approx 1/4$ resulting from the linear distributions. The interface P\'eclet number is defined by $\PeNumber_s = Uw/D_s$, and the Biot number by $\BiNumber = c_0D \omega \DaNumber /(U c_k) = \kappa_d w/U$. 

The P\'eclet number $\PeNumber_s$ is a measure of characteristic advection to diffusion rates of the interfacial surfactants, which with eq.~\eqref{eq:linThetaValue} can be written as
\begin{equation}
    r_\mathrm{adv} = Uc_k/w \quad \text{ and } \quad r_\mathrm{diff} = D_sc_k/w^2, 
    \label{eq:interfAdvDiffRates}
\end{equation}
respectively. The characteristic adsorption/desorption fluxes \eqref{eq:bulkAdsorptionRate} correspond to a rate $\kappa_d c_k$. Hence, the Biot number expresses the adsorption/desorption rate to the advection rate of interfacial surfactants. 

For a specific value of $\Delta \theta/c_k$, eq.~\eqref{eq:interfaceVel2} implies that for $u_s^0/U$ to be close to zero, the characteristic advection of interfacial surfactants must dominate (i) diffusion of interfacial surfactants ($\PeNumber_s \gg \Delta \theta/c_k$) and (ii) bulk exchange (cf.~fig.~\ref{fig:surfactantTransport}). The bulk exchange may be limited by either adsorption/desorption rate ($\DaNumber_\delta \ll 1$) or diffusion of bulk surfactants ($\DaNumber_\delta \gg 1$). If $\DaNumber_\delta \ll 1$, we must have $\BiNumber \ll \Delta \theta/c_k$, i.e. advection of interfacial surfactants must dominate over the adsorption/desorption rate. If $\DaNumber_\delta \gg 1$, then we must have $\BiNumber/\DaNumber_\delta \ll \Delta \theta/c_k$: advection of interfacial surfactants must dominate over the diffusion of bulk surfactants. 

The Marangoni stress also dictates the velocity on the interface (eq.~\ref{eq:middleInterfaceVel}). This relationship and eq.~\eqref{eq:interfaceVel2} give a condition for $\Delta \theta$, derived in the next section.

\subsection{Corresponding slip length}
\label{sec:correspondingSlipLength}
Linearisation of the Marangoni stresses with surface tension \eqref{eq:surfaceTension} implies
\begin{equation}
    \tau_\MaNumber = -\frac{\dif \gamma}{\dif x} = -\frac{\dif \gamma}{\dif \theta}\frac{\dif \theta}{\dif x} \approx \frac{nRT}{\omega}\frac{\dif \theta}{\dif x} = \mu_\infty U\MaNumber\frac{\dif \theta}{\dif x},
    \label{eq:marangoniSimpleFrumkin}
\end{equation}
where we have introduced the Marangoni number $\MaNumber = nRT/(\omega \mu_\infty U)$. We have chosen to neglect thermal effects ($T$ is a constant). The gradient of the surface coverage $\dif \theta/\dif x$ is generally positive (fig.~\ref{fig:analyticalModelAssumptions}), meaning that $\dif \gamma/\dif x$ is negative, and the Marangoni stresses act in the negative streamwise direction, as shown by fig.~\ref{fig:shearBalance}. By assuming linear interfacial concentrations, $\dif \theta/\dif x$ can be estimated as $2\Delta \theta/w$, resulting in
\begin{equation}
    \tau_\MaNumber \approx 2\mu_\infty U\MaNumber\frac{\Delta \theta}{w}.
    \label{eq:marangoniSimpleFrumkinEstimation}
\end{equation}
It follows from eq.~\eqref{eq:middleInterfaceVel} that
\begin{equation}
    u_{s}^0 \approx u_{s,\mathrm{SHS}}^0\beta_\mathrm{LIS}\left(1 - 2\MaNumber\Delta \theta \right).
    \label{eq:interfaceVel}
\end{equation}

Eq.~\eqref{eq:marangoniSimpleFrumkinEstimation} implies a scaling $nRT/(\omega w)$ of the Marangoni stress, neglecting $\Delta \theta$. Since $\tau_\infty = \mu_\infty U/w$, the Marangoni number expresses the ratio of characteristic Marangoni to imposed shear stresses. If the characteristic size of $\theta$ is considered, the Marangoni number transforms to $\MaNumber c_k$. 

The velocities expressed by \eqref{eq:interfaceVel2} and \eqref{eq:interfaceVel} must be the same. This equality results in the expression
\begin{equation}
    \Delta \theta \approx \frac{ u_{s,\mathrm{SHS}}^0\beta_\mathrm{LIS}}{U}c_k\bigg/\left(c_1 \frac{1}{\PeNumber_s} + c_2\frac{\BiNumber}{1 + \DaNumber_\delta} + 2\frac{u_{s,\mathrm{SHS}}^0\beta_\mathrm{LIS}}{U}\MaNumber c_k\right).
    \label{eq:deltaThetaModel}
\end{equation}
which by eq.~\eqref{eq:marangoniSimpleFrumkinEstimation} gives $\tau_\MaNumber$. Eq.~\eqref{eq:slipRelation} then gives the slip length
\begin{equation}
    b \approx b_\mathrm{SHS}\beta_\mathrm{LIS}\left(1 - \frac{1}{\alpha_\mathrm{diff}+ \alpha_S + 1}\right).
    \label{eq:slipLengthEstimation}
\end{equation}
We have introduced
\begin{equation}
    \alpha_\mathrm{diff} = c_1\frac{1}{\PeNumber_s'}\frac{1}{2\MaNumber c_k} \quad \text{ and } \quad \alpha_S = c_2\frac{\BiNumber' }{1 + \DaNumber_\delta}\frac{1}{2\MaNumber c_k}
    \label{eq:alpha} 
\end{equation}
where $\PeNumber_s' = wu_{s,\mathrm{SHS}}^0\beta_\mathrm{LIS}/D_s$ and $\BiNumber' = \kappa_d w /(u_{s,\mathrm{SHS}}^0\beta_\mathrm{LIS}$) are the P\'eclet and Biot numbers, respectively, with more appropriate velocity scales. 
The two expressions of \eqref{eq:alpha} represent the effects of interfacial surfactant diffusion and bulk exchange on the slip length, respectively. In order to reduce $b$, they must both be small, $\alpha_\mathrm{diff} \lesssim 1$ and $\alpha_S \lesssim 1$. Based on the interpretations of the non-dimensional numbers,
\begin{align}
    \alpha_\mathrm{diff} &\sim \frac{\text{diffusion rate of inter.~surf.}}{\text{advection rate of inter.~surf.}}\frac{\text{imposed shear stress}}{\text{Marangoni stress scale}}, \label{eq:alphaDiffMeaning} \\
    \alpha_{S} &\sim \frac{\frac{\text{adsorption/desorption rate}}{\text{advection rate of inter.~surf.}}}{1 + \frac{\text{adsorption/desorption rate}}{\text{diffusion rate of bulk surf.}}}\frac{\text{imposed shear stress}}{\text{Marangoni stress scale}}. \label{eq:alphaSMeaning} 
\end{align}
Going back to fig.~\ref{fig:frumkinSDSComparison}, we can plot the results also from the analytical model, showing a satisfactory agreement with the simulation results. Some central non-dimensional numbers are summarised in tab.~\ref{tab:nonDimNumbers}. 

\begin{table}
    \centering
    \begin{tabular}{lllllll}
    Surfactant                      & SDS             & SDS                  & SDS                & SDS          & C$_{12}$TAB  & SDS           \\ 
    Comment                         & water-air       & $\kappa_d = 500$ 1/s & $\kappa_d = 5$ 1/s & adv.~Frumkin & adv.~Frumkin & high $\tau_\infty$  \vspace{0.2cm} \\
    $u_{s,\mathrm{SHS}}/U$          & 0.31            & 0.31                 & 0.31               & 0.31         & 0.31         & 0.31 \\
    $\beta_\mathrm{LIS}$            & 0.97            & 0.19                 & 0.19               & 0.19         & 0.50         & 0.19 \\
    $\PeNumber$                     & 4.7             & 4.7                  & 4.7                & 4.7          & 4.7          & 4700 \\ 
    $\delta/w$                      & 0.28            & 0.28                 & 0.28               & 0.28         & 0.28         & 0.068 \\ 
    $\MaNumber/10^5$                & 5.8             & 2.6                  & 2.6                & 3.7          & 1.8          & 0.0037 \\ 
    $\PeNumber_s'$                  & 1.4             & 0.28                 & 0.28               & 0.28         & 0.74         & 280 \\
    $\BiNumber'/10^2$               & 50              & 250                  & 2.5                & 2.5          & 0.97        & 0.0025 \\ 
    $\DaNumber_\delta$              & 14              & 97                   & 0.97               & 0.91         & 29           & 0.22 \\
    $\alpha_S/\alpha_\mathrm{diff}$ & 59              & 9.1                  & 4.5                & 4.7          & 0.30         & 7.3 \\
    $\MaNumber^a/10^5$              & --              & --                   & --                 & 2.1          & 2.1          & 0.0021 \\ 
    \end{tabular}
    \caption{Non-dimensional numbers corresponding to the 
    results presented in secs.~\ref{sec:regularFrumkin}, \ref{sec:advancedFrumkin}, and \ref{sec:stagnantCap}. }
    \label{tab:nonDimNumbers}
\end{table}

If $\alpha_S \gg \alpha_\mathrm{diff}$, the diffusion of interfacial surfactants is insignificant compared to the bulk exchange. In the opposite situation, $\alpha_S \ll \alpha_\mathrm{diff}$, the interfacial diffusion dominates (insoluble limit). For the $\kappa_d = 500$ 1/s water-dodecane and water-air systems, the bulk exchange is much faster than interfacial diffusion ($\alpha_S \gg \alpha_\mathrm{diff}$). The parameter $\alpha_S$ thereby governs the decrease of slip length for these systems. The values of $\DaNumber_\delta$ are $97$ and $14$, respectively. Since $\DaNumber_\delta \gg 1$, the bulk exchange rate is limited by the diffusion of bulk surfactants, implying that $\alpha_S$ depends on $\kappa_d/\kappa_a$ ($\BiNumber'/\DaNumber_\delta$) instead of $\kappa_d$ ($\BiNumber'$). For the water-dodecane system, $\kappa_d = 5$ 1/s corresponds to $\DaNumber_\delta = 0.97$, below which the condition \eqref{eq:slipLengthEstimation} becomes more strict as then $\alpha_S$ becomes approximately proportional to $\kappa_d$.

\begin{figure}
  \centering 
  \includegraphics[height=5.5cm]{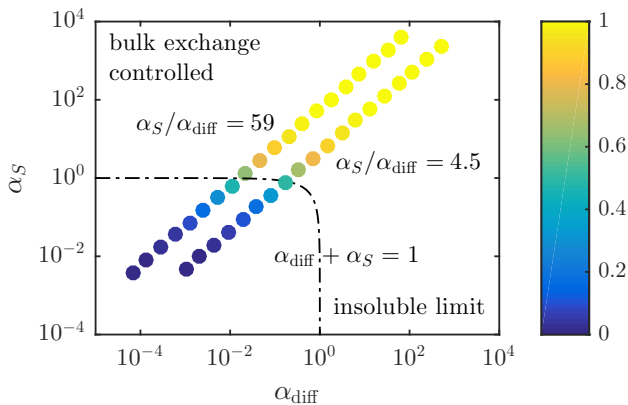} 
  \caption{ The normalised slip length $b/(b_\mathrm{SHS}\beta_\mathrm{LIS})$ of the SDS water-air and the SDS $\kappa_d = 5$ 1/s water-dodecane systems ($\alpha_S/\alpha_\mathrm{diff} = 59$ and $4.5$, respectively), plotted in the space spanned by $\alpha_\mathrm{diff}$ and $\alpha_S$. The dashed-dotted line shows $\alpha_\mathrm{diff} + \alpha_S = 1$. }
  \label{fig:simpleFrumkinSDSAlphaMap}
\end{figure}

Fig.~\ref{fig:simpleFrumkinSDSAlphaMap} illustrates how $\alpha_\mathrm{diff}$ and $\alpha_S$ can be used to predict whether there will be a considerable decrease in slip length. These two parameters span a two-dimensional space. If 
\begin{equation}
    \alpha_\mathrm{diff} + \alpha_S = 1,
    \label{eq:alphaLimit}
\end{equation} 
the slip length has halved compared to surfactant-free interfaces ($b/(b_\mathrm{SHS}\beta_\mathrm{LIS}) \approx 0.5$). In the region bounded by eq.~\eqref{eq:alphaLimit}, the slip length decrease is larger and outside it is  smaller. We use this limit to denote a significant slip length reduction, but other threshold values could also be used. Since we only varied $c_0$, $\alpha_S/\alpha_\mathrm{diff}$ is constant, describing a straight line in the $(\alpha_\mathrm{diff},\alpha_S)$ space, shown in fig.~\ref{fig:simpleFrumkinSDSAlphaMap} for the water-air and the $\kappa_d = 5$ 1/s water-dodecane systems (cf.~fig.~\ref{fig:frumkinSDSResults}). The water-dodecane system has $\alpha_S$ and $\alpha_\mathrm{diff}$ of similar magnitude ($\alpha_S/\alpha_\mathrm{diff} = 4.5$). Bulk exchange is more prominent than interfacial diffusion, but both are considerable. For the air-water system, bulk exchange dominates and the points are shifted towards the upper left corner of the figure.

\section{Adsorption and desorption taking into account alkane-surfactant interaction}
\label{sec:advancedFrumkin}
The bulk exchange model presented in sec.~\ref{sec:regularFrumkin} resulted in a more considerable surface tension decrease for the LIS than the SHS for the same SDS concentration. However, it cannot capture the initial decrease in surface tension at low concentrations appearing in water-alkane systems. To be able to describe this phenomenon, more advanced models are needed. 

The water-alkane interface abnormalities must be caused by interactions between adsorbed surfactants and alkane molecules adjacent to the interface. 
The level of interaction between adsorbed surfactant and alkane molecules at the interface can be modelled by an effective alkane interface concentration $\Gamma^a$ \citep{fainerman19}. It is associated with a molar area $\omega^a \sim \omega$ and surface coverage $\theta^a = \omega^a\Gamma^a$.  This interaction has recently been investigated on molecular levels \citep{kartashynska20, muller21}. In this manuscript, we adopt a model consistent with the equilibrium model of \citet{fainerman19}. They assumed that the molar area $\omega$ decreased with surface coverage. However, this decrease is only notable for higher concentrations ($\theta \sim 1$), and, therefore, we neglect this correction.

The interfacial and bulk surfactant concentrations are assumed to follow the same transport equations as the previous model (eqs.~\ref{eq:interfacialTransport} and \ref{eq:bulkTransport}, respectively). However, the source term is (cf.~\ref{eq:simpleSource})
\begin{equation}
    S = \frac{1}{\omega}\left(\kappa_{a} c_s\left(1 - \theta^t\right) - \kappa_{d}\theta e^{-2a\theta - 2a^{as}\theta^a}\right),
    \label{eq:advSource}
\end{equation}
where we have introduced the additional interaction constant $a^{as}$ and the total surface coverage $\theta^t = \theta + \theta^a$. The adsorption term contains the total surface coverage, but the desorption term maintains its $\theta$-dependency (together with an additional exponential factor). 

We assume that the alkane molecules are in local equilibrium with the adsorbed surfactants, resulting in the corresponding expression 
\begin{equation}
    \kappa_a^a c_0^a\left(1 - \theta^t\right) - \kappa_d^a\theta^a e^{-2a^a\theta^a - 2a^{as}\theta} = 0,
    \label{eq:alkaneSource}
\end{equation}
where $\kappa_a^a$ and $\kappa_d^a$ are the adsorption and desorption coefficients of the alkane phase, respectively, $a^a$ is a constant, and $c_0^a$ is the (constant) bulk alkane concentration. The current model assumes that the alkane adsorption coefficient depends on $\theta$ by 
\begin{equation}
    \kappa_a^a = \min(\kappa_{a,0}^{a}\theta, \kappa_{a,\mathrm{max}}^a),
    \label{eq:alkaneAdsorptionCoefficient}
\end{equation}
where $\kappa_{a,0}^{a}$ and $\kappa_{a,\mathrm{max}}^a$ are constants. This expression implies that the alkane adsorption rate increases proportionally to $\theta$ for minuscule concentrations. Without surfactants, $\theta^a = 0$.

The surface tension has been modelled by (cf.~\ref{eq:surfaceTension})
\begin{equation}
    \gamma = \gamma_c + \frac{RT}{\omega_0}\left[\ln(1 - \theta^t) + a\theta^2 + a^a{\theta^a}^2 + 2a^{as}\theta \theta^a\right],
    \label{eq:advSurfaceTension}
\end{equation}
where
\begin{equation}
    \omega_0 = \frac{1}{\theta^t}\left(\frac{\omega}{n}\theta + \omega^a\theta^a\right)
\end{equation}
is the effective average molar area.



\begin{figure}
  \centering
     \begin{subfigure}{1.0\textwidth}
      \centering
      \includegraphics[height=4.5cm]{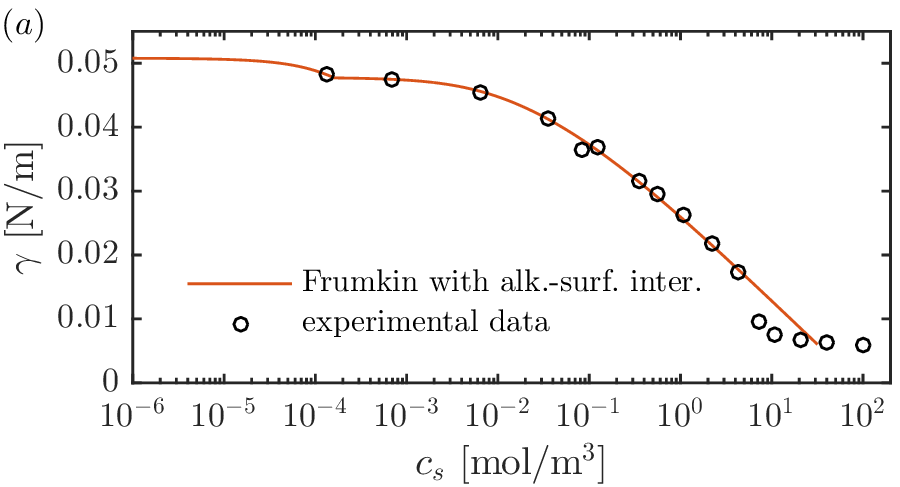} 
       \captionlistentry{}
      \label{fig:equiSurfTenC12TABExp}
  \end{subfigure}
  \begin{subfigure}{0.48\textwidth}
      \centering
      \includegraphics[height=5.5cm]{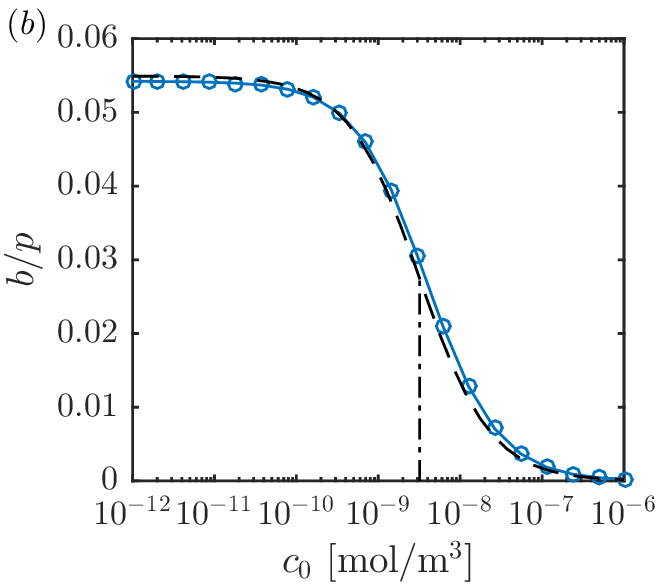} 
      \captionlistentry{}
      \label{fig:frumkinC12TABComparison}
  \end{subfigure}
  \begin{subfigure}{0.48\textwidth}
      \centering
      \includegraphics[height=5.5cm]{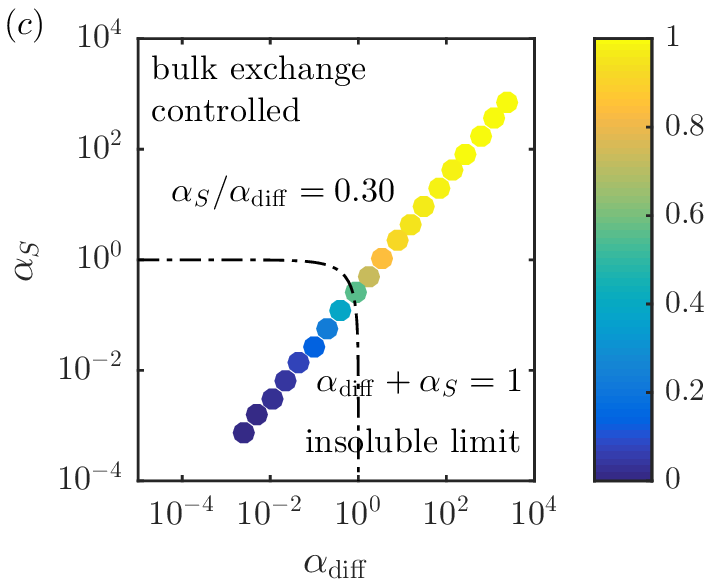} 
      \captionlistentry{}
      \label{fig:frumkinC12TABAlphaMap}
  \end{subfigure}
  \caption{ ($a$) Equilibrium surface tension of C$_{12}$TAB at a water-hexane interface with experimental data from \citet{pradines10}. Without surfactants, the surface tension is $\gamma_c = 50.8$ mN/m for water-hexane (20\degree C, \citealt{zeppieri01}). ($b$) Slip lengths at $\tau_\infty = 0.33$~mPa for different bulk concentrations of C$_{12}$TAB at a water-hexane interface, together with the analytical model (dashed lines). The correspondence for SDS is shown in fig.~\ref{fig:frumkinSDSComparison}. ($c$) The normalised slip length, $b/(b_\mathrm{SHS}\beta_\mathrm{LIS})$, plotted in the parameter space of $\alpha_\mathrm{diff}$ and $\alpha_S$. In both ($b$) and ($c$), the dashed-dotted lines are $\alpha_\mathrm{diff} + \alpha_S = 1$, corresponding to $b/(b_\mathrm{SHS}\beta_\mathrm{LIS}) \approx 1/2$ (eq.~\ref{eq:slipLengthEstimation}). }
  \label{fig:frumkinC12TABResults}
\end{figure}

The results plotted in fig.~\ref{fig:frumkinSDSResults} show that this improved modelling lowers the critical concentration by an additional order of magnitude. Corresponding results for C$_{12}$TAB and water-hexane LIS are shown in fig.~\ref{fig:frumkinC12TABResults}. We only used the more advanced adsorption/desorption model for C$_{12}$TAB (parameters given in tab.~\ref{tab:surfactantProperties}). This system manifests a stronger sensitivity to the surfactants, as $b/p$ decreases significantly at even lower concentrations. The simulations are illustrated in the $(\alpha_\mathrm{diff},\alpha_S)$ space in fig.~\ref{fig:frumkinC12TABAlphaMap}. In contrast to SDS (fig.~\ref{fig:simpleFrumkinSDSAlphaMap}), C$_{12}$TAB have more prominent interfacial diffusion than bulk exchange ($\alpha_S/\alpha_\mathrm{diff} = 0.30$). Therefore, the simulations are closer to the bottom right corner of the figure.

\subsection{Analytical model with alkane-surfactant interaction}
Even if the source term has been changed slightly, the analytical model developed in sec.~\ref{sec:analyticalSlipModel} is essentially the same. The linear approximation of eq.~\eqref{eq:advSource} is equal to \eqref{eq:simpleSourceNonDimlinearised}. With the surfactant transport equations unchanged, the bulk exchange and the interfacial transport predictions do not need to be modified (secs.~\ref{sec:modellingBulkExchange} and \ref{sec:surfactantTransportBalance}, respectively).

By linearising the alkane source term \eqref{eq:alkaneSource}, an estimation of $\theta^a$ can be found. We introduce the non-dimensional alkane concentrations $c_{k,0}^a = c_0^a\kappa_0^a/\kappa_{d}^a$ 
and $c_{k,\mathrm{max}}^a = c_0^a\kappa_{a,\mathrm{max}}^a/\kappa_{d}^a$.
For large concentrations, $\theta^a \approx c_{k,\mathrm{max}}^a$, whereas for small concentrations, $\theta^a \approx c_{k,0}^a\theta \approx c_{k,0}^a c_k$ (eq.~\ref{eq:alkaneAdsorptionCoefficient}). Capturing both cases,
\begin{equation}
    \theta^a \approx \min(c_{k,0}^a \theta, c_{k,\mathrm{max}}^a) \approx \min(c_{k,0}^a c_k, c_{k,\mathrm{max}}^a) 
    \quad \text{ at } \quad x = x_0.
    \label{eq:linThetaAlkaneValue}
\end{equation}
All simulations have resulted in values in the lower concentration interval ($c_{k,0}^a \theta \le c_{k,\mathrm{max}}^a$). 
It follows that $\Delta \theta \approx \Delta \theta^a/c_{k,0}^a$.

We use the linear approximation of the surface tension \eqref{eq:advSurfaceTension} to estimate the Marangoni stresses. With $c_{k,0}^a > 1$ (tab.~\ref{tab:surfactantProperties}), $\theta^a$ is assumed to be larger than $\theta$ at low concentrations (eq.~\ref{eq:linThetaAlkaneValue}). With $\omega_0 \approx \omega^a$,
\begin{equation}
    \frac{\dif \gamma}{\dif x} = \frac{\partial \gamma}{\partial \theta}\frac{\dif \theta}{\dif x} + \frac{\partial \gamma}{\partial \theta^a}\frac{\dif \theta^a}{\dif x} \approx -\frac{RT}{\omega^a}\left(\frac{\dif \theta}{\dif x} + \frac{\dif \theta^a}{\dif x}\right) \approx -\mu_\infty U\MaNumber^a\frac{\dif \theta^a}{\dif x},
    \label{eq:linearisedMarangoniFrumkinFrumkin}
\end{equation}
where $\MaNumber^a = RT/(\omega^a \mu_\infty U)$ (cf.~\ref{eq:marangoniSimpleFrumkin}). Analogous to \eqref{eq:marangoniSimpleFrumkinEstimation}, $\dif \theta^a/\dif x \approx 2\Delta \theta^a/w \approx 2c_{k,0}^a\Delta \theta/w$, giving
\begin{equation}
    \tau_\MaNumber \approx 2\mu_\infty U\MaNumber^a c_{k,0}^a\frac{\Delta \theta}{w}.
    \label{eq:marangoniAdvFrumkinEstimation}
\end{equation}

We now adapt the estimation of the slip length \eqref{eq:slipLengthEstimation}. Eqs.~ \eqref{eq:middleInterfaceVel}, \eqref{eq:interfaceVel2}, and \eqref{eq:marangoniAdvFrumkinEstimation} give an expression for $\Delta \theta$ corresponding to eq.~\eqref{eq:deltaThetaModel}. The slip length is found from eq.~\eqref{eq:slipRelation} and is identical to \eqref{eq:slipLengthEstimation} if $\MaNumber$ is replaced by $\MaNumber^ac_{k,0}^a$, equivalent to redefining
\begin{equation}
    \alpha_\mathrm{diff} = c_1\frac{1}{\PeNumber_s'}\frac{1}{2\MaNumber^a c_{k,0}^a c_k} \quad \text{ and } \quad \alpha_S = c_2\frac{\BiNumber' }{1 + \DaNumber_\delta}\frac{1}{2\MaNumber^a c_{k,0}^a c_k}.
    \label{eq:alphaFrumkinFrumkin}
\end{equation}
The results from this analytical model are also shown in figs.~\ref{fig:frumkinSDSResults} and \ref{fig:frumkinC12TABComparison}, in good agreement with the simulation results.

The sudden change in the gradient of $\gamma$ in figs.~\ref{fig:equiSurfTenSDSExp} and \ref{fig:equiSurfTenC12TABExp} ($c_s = 1.1\cdot10^{-2}$ and $1.6\cdot10^{-4}$ mol/m$^3$, respectively) coincide with switching the $\theta$-dependency of $\kappa_a^a$ in eq.~\eqref{eq:alkaneAdsorptionCoefficient}. However, the simulations are unaffected since they are performed at lower concentrations. In this interval, $\partial \gamma/\partial \theta^a \approx -RT/\omega^a$ (eq.~\ref{eq:linearisedMarangoniFrumkinFrumkin}), which is independent of concentration. Since $\theta^a$ depends linearly on $\theta$ (eq.~\ref{eq:linThetaAlkaneValue}), the total derivative with respect to $\theta$ is constant, and with respect to $c_0$ in equilibrium since $\theta \approx c_k = \kappa_ac_0/\kappa_d$ (eq.~\ref{eq:linThetaValue}). This concentration interval has no experimental data to validate the surface tension curve. 
A linear decrease in surface tension with concentration (i.e.~constant derivative) is a reasonable first assumption. 

Another problem of the current adsorption/desorption model is that since $c_{k,0}^a$ usually is large, the ratio between $\Gamma$ and $\Gamma^a$ becomes unreasonably large at low concentrations, considering the number of alkane molecules that can interact with one surfactant molecule. A more realistic result is achieved by letting the surfactant molecules adsorb in two different states, described by the so-called reorientation model \citep{kartashynska20}. We welcome future studies using more advanced adsorption/desorption models. However, for this investigation, we consider the current model sufficient.

\section{Stagnant cap regime}
\label{sec:stagnantCap}
High shear stresses can result in a portion of the interface with almost no adsorbed surfactants. In such conditions, the interface is no longer in the uniformly retarded regime -- an underlying assumption of the analytical model. The downstream part of the interface with surfactants becomes a stagnant cap (SC) in analogy to bubbles \citep{palaparthi06}. The interface can be said to be in a partial SC regime. A slip length is partly regained since the flow is not decelerated outside the stagnant part. If the concentration of bulk surfactants cannot be reduced further, a partial SC regime can therefore be desirable to achieve. The partial SC regime was investigated by \citet{baier21} for insoluble surfactants on SHS in the limit of large $\PeNumber_s'$. For LIS, the stagnant cap can improve the retention of the oil in the groove, but on the other hand, the stagnant part of the interface does not contribute to drag reduction \citep{fu19}.

This section shows that a SC eventually grows to cover the whole interface if the bulk concentration increases. The interface is then in the so-called full SC regime, again undesirable for drag reduction. There is no apparent difference in the distribution of $\theta$ between the full SC and the uniformly retarded regime, so distinguishing between them is a mere formality. 

To form a SC, we increased the imposed shear stress to $\tau_\infty = 0.33$ Pa (factor of 1000) by increasing the applied velocity gradient and thereby $U$. Relevant affected non-dimensional parameters are $\PeNumber$, $\PeNumber_s'$, $\BiNumber'$, and $\MaNumber$, of which the first two increase linearly with $U$ and the latter decrease linearly (tab.~\ref{tab:nonDimNumbers}). The increase in $\PeNumber$ makes the concentration boundary layer thinner (eq.~\ref{eq:delta}). Interestingly, $\alpha_\mathrm{diff}$ and $\alpha_S$ are independent of $U$ in the Stokes regime, except through $\delta$. 
The Reynolds number is $\ReNumber = \rho_\infty Uw/\mu_\infty = 3.3 \sim 1$ (with density $\rho_\infty = 1000$ kg/m$^3$ for water), 
so the Stokes equations are still valid to some extent, especially considering the lower velocity close to the interfaces.

A partial SC requires that the advection rate of interfacial surfactants overcomes the interfacial diffusion and bulk exchange rates of the linear surfactant distribution. 
The non-dimensional numbers comparing these transport mechanisms are reiterated here for convenience, (cf.~\ref{eq:alphaDiffMeaning} and \ref{eq:alphaSMeaning})
\begin{align}
    &\frac{\text{diffusion rate of inter.~surf.}}{\text{advection rate of inter.~surf.}} \sim \frac{1}{\PeNumber_s'}, \\
    &\frac{\text{diffusion rate of bulk surf.}}{\text{advection rate of inter.~surf.}} \sim \frac{\BiNumber'}{\DaNumber_\delta}, \\ 
    &\frac{\text{adsorption/desorption rate}}{\text{advection rate of inter.~surf.}} \sim \BiNumber'. 
\end{align} 

The quantitative limits of the SC formation can be understood from eq.~\eqref{eq:deltaThetaModel}. This equation can be reformulated as
\begin{equation}
    1 - \frac{1}{\alpha_\mathrm{diff}+ \alpha_S + 1} \approx \left(c_1 \frac{1}{\PeNumber_s'} + c_2\frac{\BiNumber'}{1 + \DaNumber_\delta}\right)\frac{\Delta \theta}{c_k}.
    \label{eq:interfaceVelCorrectScaling}
\end{equation}
A linear interfacial surfactant distribution implies $\Delta \theta/c_k \le 1$ since we cannot have negative values of $\theta$ (fig.~\ref{fig:analyticalModelAssumptions}). Therefore, eq~\eqref{eq:interfaceVelCorrectScaling} cannot hold for large $\alpha_\mathrm{diff}+ \alpha_S$ if
\begin{equation}
    C_\mathrm{SC} = c_1 \frac{1}{\PeNumber_s'} + c_2\frac{\BiNumber'}{1 + \DaNumber_\delta} < 1.
    \label{eq:requirementForExistenceOfPSC}
\end{equation}
Instead, the surfactants are advected to the end stagnation point of the interface, where a larger gradient is created. The interface is then in the partial SC regime. The same limitation exists for $\Delta c_s/c_0$. However, from eq.~\eqref{eq:cGradScaling}, we know that $\Delta c_s/c_0 \lesssim \Delta \theta/c_k$, so it does not provide any other peculiarities. If $\alpha_\mathrm{diff} + \alpha_S \ll 1$, \eqref{eq:interfaceVelCorrectScaling} would again allow for a linear surfactant profile. Such interface is in the full SC regime. We suggest using eq.~\eqref{eq:requirementForExistenceOfPSC} to distinguish between (i) the uniformly retarded regime ($C_\mathrm{SC} > 1$) and (ii) the partial and full SC regimes ($C_\mathrm{SC} < 1$).

\begin{figure}
  \centering
  \begin{subfigure}{0.48\textwidth}
      \centering
      \includegraphics[height=5cm]{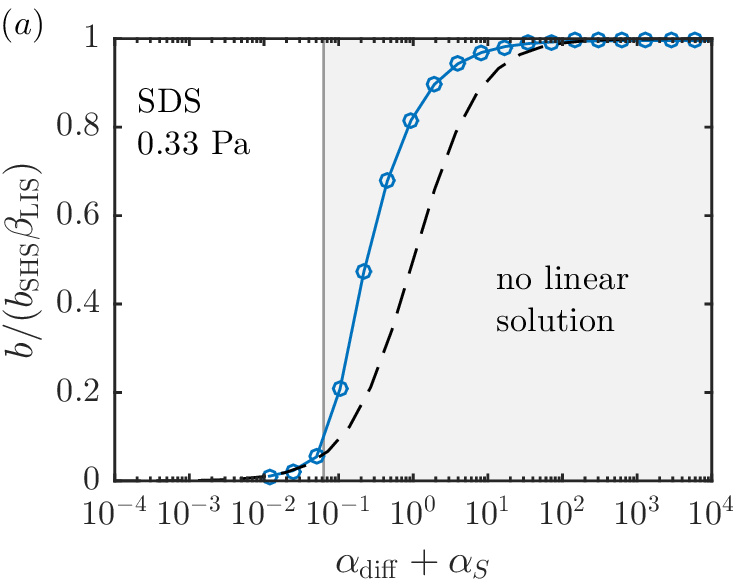} 
       \captionlistentry{}
       \label{fig:033SDSAlphaMap}
  \end{subfigure}
  \begin{subfigure}{0.48\textwidth}
      \centering
      \includegraphics[height=5cm]{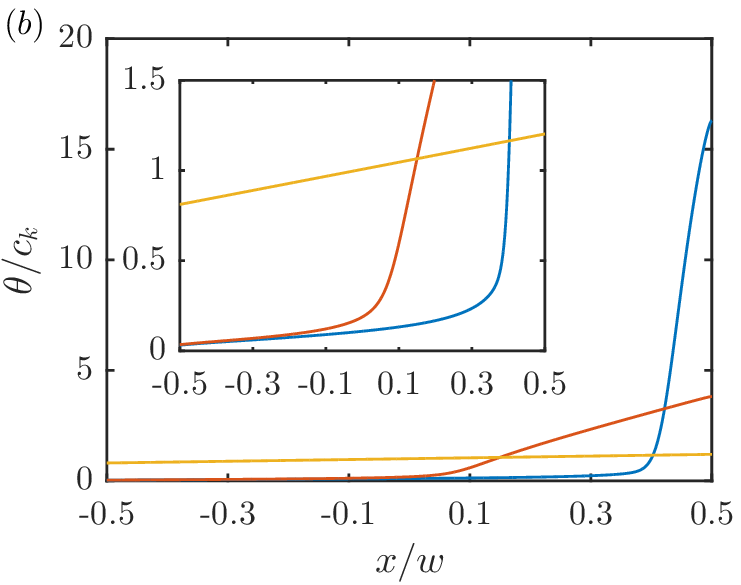} 
       \captionlistentry{}
  \end{subfigure}
  \caption{($a$) The analytical model (dashed line) compared to simulations of SDS at $\tau_\infty = 0.33$ Pa. The limit \eqref{eq:linearLimit} forms the boundary of the grey region, in which linear distributions of $\theta$ cannot exist. ($b$) Three distributions of $\theta/c_k$, going from a non-linear (blue) to an intermediate (red) and a linear (yellow) for the smallest value of $\alpha_\mathrm{diff} + \alpha_S$. The inset is identical, apart from a lower maximum vertical axis limit.}
  \label{fig:033SDS}
\end{figure}

Assuming that eq.~\eqref{eq:requirementForExistenceOfPSC} is fulfilled, and considering the maximum value $\Delta \theta/c_k = 1$, we get the limit for $\alpha_\mathrm{diff}+ \alpha_S$ above which no linear solution can exist as
\begin{equation}
    \alpha_\mathrm{diff} + \alpha_S \approx \frac{1}{1 - C_\mathrm{SC}} - 1.
    \label{eq:linearLimit}
\end{equation}
Eq.~\eqref{eq:linearLimit} is the lower limit for the partial SC regime, below which we enter the full SC regime. 

We illustrate the limit \eqref{eq:linearLimit} for SDS (advanced model) at $\tau_\infty = 0.33$ Pa in fig.~\ref{fig:033SDS}. These simulations were performed with a more refined grid (appendix \ref{sec:gridStudy}). The region where linear solutions cannot exist is shown in grey. For the current parameters (tab.~\ref{tab:nonDimNumbers}),
\begin{equation}
    \frac{1}{\PeNumber_s'} = 0.0036, \quad \BiNumber' = 0.25, \quad \text{ and } \quad \frac{\BiNumber'}{\DaNumber_\delta} = 1.2, \quad \text{giving} \quad C_\mathrm{SC} = 0.059.
    \label{eq:currentParameters}
\end{equation}
In the grey region, the analytical model loses validity as expected. Eq.~\eqref{eq:linearLimit} also holds for the simulation results of \citet{landel20}, which are shown in appendix \ref{sec:landel2020}. Therefore, the transition between the partial and full SC regime appears consistent over a wide range of geometrical and surfactant parameters.

The limit for the bulk surfactant concentration corresponding to \eqref{eq:linearLimit} is attained by using the definitions of $\alpha_\mathrm{diff}$ and $\alpha_S$ (eqs.~\ref{eq:alpha} or \ref{eq:alphaFrumkinFrumkin}). The expression for the non-dimensional concentration is
\begin{equation}
    c_k = \frac{1}{2\MaNumber}\left(1 - C_\mathrm{SC}\right) \quad \text{ or } \quad c_k^a = \frac{1}{2\MaNumber^a}\left(1 - C_\mathrm{SC}\right),
    \label{eq:linearConcentrationLimit}
\end{equation}
for the regular and the more advanced Frumkin model, respectively, where $c_k^a = c_{k,0}^a c_k$. No partial SC can exist above these concentrations. As $C_\mathrm{SC}$ decreases, $c_k$ or $c_k^a$ increases towards a maximum value determined by the Marangoni number. The current parameters (eq.~\ref{eq:currentParameters}) correspond to $94$\% of the maximum concentration. The only flow-dependent quantity of the maximum concentration is $U$, which is determined by the imposed shear stress and the width of the grooves. Apart from $T$, the other quantities depend only on the liquids and the surfactants.

With the increase of slip length in the partial SC regime, a new condition for significant slip length reduction is needed in place of eq.~\eqref{eq:alphaLimit}. Eq.~\eqref{eq:linearLimit} can replace eq.~\eqref{eq:alphaLimit} if smaller than $1$ -- this occurs if $C_\mathrm{SC} < 1/2$ -- since \eqref{eq:alphaLimit} still is valid in the full SC regime. However, it should be noted that eq.~\eqref{eq:linearLimit} then gives $b/b_\mathrm{SHS}\beta_\mathrm{LIS} < 0.5$ as predicted by the analytical model becoming valid at this $\alpha_\mathrm{diff} + \alpha_S$. Still, we do think this is an appropriate condition. We summarise these limits by concluding that a significant slip length reduction occurs if
\begin{equation}
    \alpha_\mathrm{diff} + \alpha_S <
    \begin{cases}
        1   & \text{if } C_\mathrm{SC} \ge 1/2, \\
        \dfrac{1}{1 - C_\mathrm{SC}} - 1 & \text{if } C_\mathrm{SC} < 1/2.
    \end{cases} 
    \label{eq:alphaLimitGeneral}
\end{equation}

The equations provided in this section are intended to serve as convenient tools to improve the understanding and classification of LIS with surfactants subjected to high shear stresses. We have not attempted to model the partial SC regime, which is a potential topic for future studies. However, it can be noted that, even in the partial SC regime, the analytical model can indicate the order of magnitude of the slip length. Uncertainties in parameters and adsorption models might be a larger concern; the difference between the analytical prediction and the simulation results of fig.~\ref{fig:033SDSAlphaMap} can be compared to the deviations in slip length in fig.~\ref{fig:frumkinSDSResults}. The error of the analytical model can, in particular, be expected to be minor if $C_\mathrm{SC} \sim 1$, implying that the two conditions for significant slip length reduction (\ref{eq:alphaLimit} and \ref{eq:alphaLimitGeneral}) are similar.

\section{Remarks}
\label{sec:remarks}

The analytical model has been developed for two-dimensional transverse grooves. However, similar surfactant transport processes are expected to be present for LIS with three-dimensional longitudinal grooves \citep{landel20, temprano21}. 
Therefore, the analytical model is expected to provide an indication of the effects of surfactants on the slip length also for such configurations, commonly used in experimental LIS studies \citep{wexler15b, jacobi15, fu19}. 

If the grooves are longer than those used here, the slip length tends to be larger. When disregarding changes in aspect ratio or solid fraction, $b_\mathrm{SHS}\beta_\mathrm{LIS}$ and $u_{s,\mathrm{SHS}}^0\beta_\mathrm{LIS}$ increase proportional to $w$ (as given by eqs.~\ref{eq:idealSHSSlipLengthRelation} and \ref{eq:uSRelation}). 
The diffusion rate is suppressed by increasing the groove width. Thereby, $\PeNumber_s'$ 
and $\DaNumber_\delta$ 
increase with $w$ (even if the $w$-dependency of $\delta$ in eq.~\ref{eq:delta} is non-trivial), whereas $\BiNumber'$ is constant. Correspondingly, an increased groove width decreases $C_\mathrm{SC}$ (eq.~\ref{eq:requirementForExistenceOfPSC}); increasing the groove width may make the interfaces enter the SC regime. As $\MaNumber$ and $\MaNumber^a$ are inversely proportional to $w$, $\alpha_\mathrm{diff}$ decreases with $w$, whereas $\alpha_S$ increases (eqs.~\ref{eq:alpha} or \ref{eq:alphaFrumkinFrumkin}). Therefore, with $C_\mathrm{SC}$ decreasing and $\alpha_S$ eventually dominating, the slip length becomes less sensitive to surfactants with increasing groove width, according to eq.~\eqref{eq:alphaLimitGeneral}. 

For turbulent flows, the grooves would need to be larger or the imposed shear stress higher to generate a relevant drag reduction, even without surfactants. For $\tau_\infty = 0.33$ mPa and $0.33$ Pa, the viscous length scales correspond to $l_\nu = \mu_\infty/\sqrt{\rho_\infty\tau_\infty} = 1.7$ mm and $55$ \textmu m, respectively. Assuming $b/p \approx 0.02$ also for the turbulent flows (fig.~\ref{fig:frumkinSDSResults}), $b^+ = b/l_\nu \approx 1.7\cdot10^{-3}$ and $5.4\cdot10^{-2}$, but these non-dimensional slip lengths would need to be $b^+ \gtrsim 1$ to give meaningful drag reduction \citep{fu17}. 

Some surfactants exhibit a critical micelle concentration (CMC), above which they can form micelles (clusters of surfactant molecules) instead of adsorbing at interfaces. Hence, the source term dependencies on $c_s$ (eqs.~\ref{eq:simpleSource} and \ref{eq:advSource}) are invalid above the CMC. The CMC is 8.2 mol/m$^3$ for SDS, above which the surface tension remains somewhat constant (\citealt{elworthy66}, cf.~fig.~\ref{fig:equiSurfTenSDSExp}). C$_{12}$TAB holds a similar CMC (\citealt{klevens48}, cf.~fig.~\ref{fig:equiSurfTenC12TABExp}). The slip length reductions studied here occur at lower concentrations (figs.~\ref{fig:frumkinSDSComparison} and \ref{fig:frumkinC12TABComparison}). Therefore, micelle formation does not affect our conclusions. Micelles can also act as monomer buffers (surfactants not part of micelles), dissociating to reduce monomer concentration gradients. They thereby give rise to a remobilisation effect where interface gradients are reduced \citep{manikantan20}. 

For $\theta \sim c_k \gg 1$, the analytical model could be inaccurate since non-linearities in the Frumkin source term then become relevant. As discussed by \citet{landel20}, such non-dimensional concentrations could appear below the CMC for ``strong'' surfactants ($\kappa_d/\kappa_a$ much lower than the CMC, cf.~tab.~\ref{tab:surfactantProperties}). However, they performed some simulations with $c_k \gg 1$, which roughly agreed with the model (appendix \ref{sec:landel2020}). Peculiarities that might arise at high concentrations (e.g.~for $a \neq 0$) are out of the scope of the current investigation. 

We have isolated the effects of Marangoni stresses and viscous stresses imposed by the infusing liquids. However, surfactants might also cause additional rheological stresses at the interface, characterised by the (intrinsic) surface shear and dilatational viscosities \citep{manikantan20}. However, these surface viscosities are challenging to measure. Careful measurements have given a reliable upper limit of the surface shear viscosity for surfactants on water-air interfaces \citep{zell14}. If it is assumed that this upper limit holds for both surface viscosities, these extra stresses are irrelevant compared to the Marangoni stress at groove dimensions $w \gg 10$ \textmu m \citep{landel20}. They can then be neglected for the geometries considered here.

Finally, a possible extension of the current work would be to model surfactant solubility also in the oil phase. Such a model requires a second set of surfactant adsorption and desorption coefficients, describing surfactant exchange in the oil. 

\section{Conclusions}
\label{sec:conclusions}
If surfactants are present in a flow over SHS or LIS, they might severely decrease the slip lengths of these surfaces. The surfactants adsorb at the fluid interfaces and are advected towards the downstream stagnation points, resulting in surface tension gradients that oppose the flow. Surface tension gradients are known as Marangoni stresses. 
Using numerical simulations of laminar flow over LIS with transverse grooves, we have explored how the effective slip length changes when the flow has a non-zero concentration of the commonly used surfactants SDS and C$_{12}$TAB. The external fluid was water, and the infusing liquid an alkane. The surfactants have been assumed to be soluble in the water but insoluble in the alkane. 

For low applied shear stresses ($\tau_\infty = 0.33$ mPa), the distribution of surfactants on an interface is approximately linear. It is then possible to construct an analytical theory for the LIS slip length, similar to what has been done for SHS \citep{landel20}. In the numerical simulations, we used classical Frumkin kinetics and a more advanced Frumkin model that considers interactions between the surfactants and the alkane molecules. Both models can be described by the (properly adjusted) analytical theory. The predicted slip length is given by eq.~\eqref{eq:slipLengthEstimation}, with parameters $\alpha_\mathrm{diff}$ and $\alpha_S$ specified by eq.~\eqref{eq:alpha} or \eqref{eq:alphaFrumkinFrumkin}, depending on the model.

The interfacial surfactants might be swept towards the downstream stagnation point for large applied shear stresses ($\tau_\infty = 0.33$ Pa). The interfaces are then in the partial stagnant cap regime. Some slip length is regained, as a large part of an interface does not have a significant surface tension gradient. However, this regime can only exist below a particular bulk concentration. Above this concentration, the surfactant distribution is linear, and the analytical model becomes accurate. The concentration is given by eq.~\eqref{eq:linearConcentrationLimit}, with $1 - C_\mathrm{SC}$ typically having a value close to unity.

Surfactants in fluid systems are difficult to detect directly yet affect flows significantly; they act like hidden variables \citep{manikantan20}. The concentrations for which the slip is profoundly reduced for LIS and SHS can be assumed to correspond to those naturally occurring in experimental setups \citep{peaudecerf17}. \citet{temprano21} estimated such naturally occurring surfactant concentrations to be $c_0 = 3\cdot10^{-4}$ mol/m$^3$ by comparing experimental results of SHS with polydimethylsiloxane (PDMS) textures to theoretical predictions (cf.~figs.~\ref{fig:frumkinSDSResults} and \ref{fig:frumkinC12TABComparison}). It is the hope that the models presented here can be used to interpret experimental results of LIS both with and without artificially added surfactants. They can also aid the design of grooves to avoid severe performance degradation. For example, LIS can be designed to remain in the partial stagnant cap regime at similar concentrations. 
We also hope that by including infusing liquid viscosity in this model, we have taken a step towards three-dimensional models of LIS with textures more similar to those typically used in experimental setups.

\backsection[Acknowledgements]{
This work was supported by SSF, the Swedish Foundation for Strategic Research (Future Leaders grant FFL15:0001). We are thankful to U\v{g}is L\={a}cis for his help in setting up the FreeFem++ framework. }

\backsection[Declaration of interests]{The authors report no conflict of interest.}

\appendix

\section{Grid convergence studies}
\label{sec:gridStudy}
The simulations of secs.~\ref{sec:analyticalViscousInfusedLiquid}, \ref{sec:regularFrumkin}, and \ref{sec:advancedFrumkin} were performed with a mesh spacing $w/N$ with $N = 256$ at the interface. The simulations of sec.~\ref{sec:stagnantCap} with the high shear stress $\tau_\infty = 0.33$ Pa were performed with $N = 1024$. Since $w = 100$~\textmu m, $N = 1024$ corresponds to a mesh spacing of $0.098$~\textmu m.

\begin{figure}
    \centering
    \begin{subfigure}{0.45\textwidth}
        \centering
        \includegraphics[height=4.5cm]{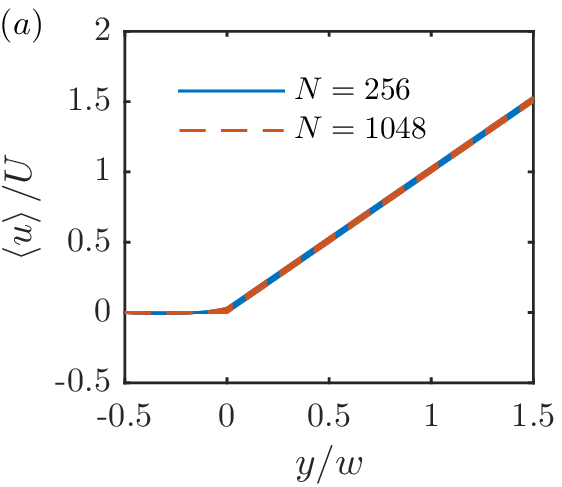} 
        \captionlistentry{}
        \label{fig:refinementStudyVel}
    \end{subfigure}
    \begin{subfigure}{0.45\textwidth}
        \centering
        \includegraphics[height=4.5cm]{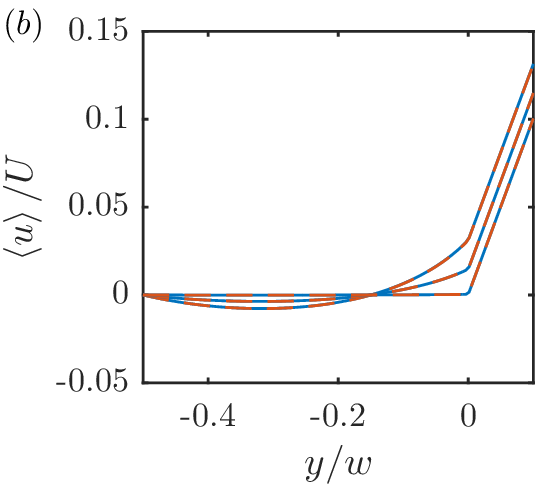} 
        \captionlistentry{}
        \label{fig:refinementStudyVelZoomIn}
    \end{subfigure}
    \begin{subfigure}{0.45\textwidth}
        \centering
        \includegraphics[height=4.5cm]{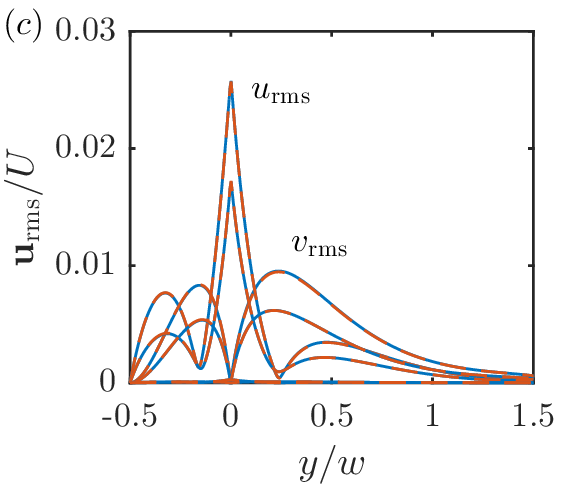} 
        \captionlistentry{}
        \label{fig:refinementStudyRmsVel}
    \end{subfigure}
    \begin{subfigure}{0.45\textwidth}
        \centering
        \includegraphics[height=4.5cm]{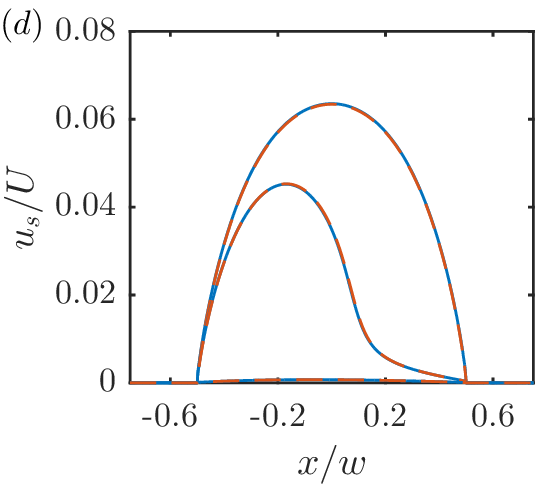} 
        \captionlistentry{}
        \label{fig:refinementStudyInterfVel}
    \end{subfigure}
    \begin{subfigure}{0.45\textwidth}
        \centering
        \includegraphics[height=4.5cm]{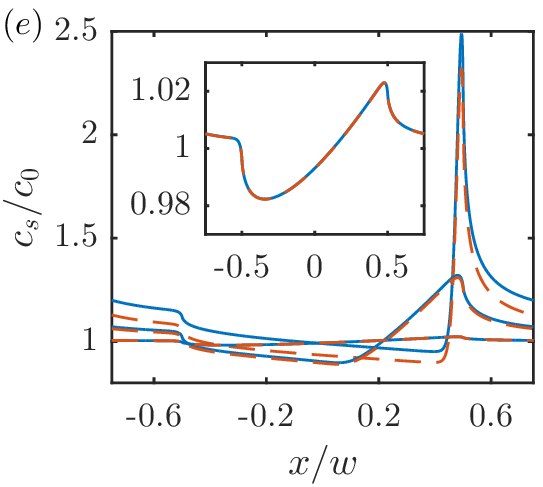} 
        \captionlistentry{}
        \label{fig:refinementStudyCs}
    \end{subfigure}
    \begin{subfigure}{0.45\textwidth}
        \centering
        \includegraphics[height=4.5cm]{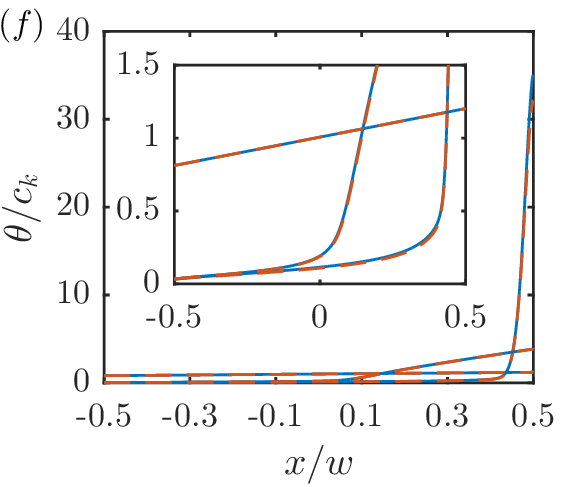} 
        \captionlistentry{}
        \label{fig:refinementStudyTheta}
    \end{subfigure}
    
    \caption{Statistics of flow with SDS for two grids and three values of $c_0$ ($\tau_\infty = 0.33$ Pa). The parameter $N$ gives the cell spacing at the interface by $w/N$. The figures illustrate: ($a$) mean velocity profiles with ($b$) a zoom-in around the interface, ($c$) root-mean-squared velocity fluctuations ($u_\mathrm{rms} = \sqrt{\mean{(u - \mean{u})^2}}$ and $v_\mathrm{rms} = \sqrt{\mean{v^2}}$ in the streamwise and wall-normal directions, respectively), ($d$) interfacial velocity, ($e$) bulk surfactant concentration at the interface, and ($f$) interfacial surfactant concentration. Velocities are normalised by $U = w\tau_\infty/\mu_\infty$ and decrease in magnitude with increasing $c_0$. The maximum values of $c_s/c_0$ and $\theta/c_k$ decrease with increasing $c_0$. Insets show zoom-ins of ($e$) $c_s/c_0$ for the highest $c_0$ and ($f$) $\theta/c_k$ for all three $c_0$.}
    \label{fig:gridRefinement}
\end{figure}

In order to test the grid dependency of the simulations, three high shear stress simulations were repeated with $N = 256$. These cases included a highly skewed $\theta$ distribution ($c_0 = 10^{-10}$ mol/m$^3$), a linear distribution ($c_0 = 10^{-4}$ mol/m$^3$), and an intermediate case ($c_0 = 5.5\cdot10^{-6}$ mol/m$^3$). Statistics are shown in fig.~\ref{fig:gridRefinement}. Overall, the results from the two grids agree well. The largest differences are found for the highly skewed distribution in $c_s$ (bulk surfactant concentration at the interface) and the peak of $\theta$, figs.~\ref{fig:refinementStudyCs} and \ref{fig:refinementStudyTheta}, respectively.

\begin{figure}
    \centering
    \begin{subfigure}{0.45\textwidth}
        \centering
        \includegraphics[height=4.5cm]{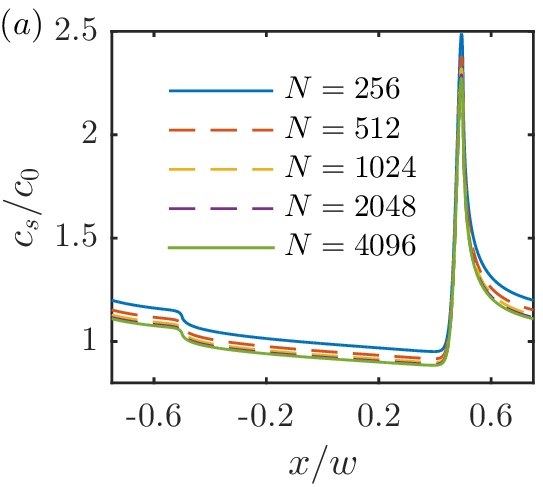} 
        \captionlistentry{}
        \label{fig:refinementStudyCsLowestC0}
    \end{subfigure}
    \begin{subfigure}{0.45\textwidth}
        \centering
        \includegraphics[height=4.5cm]{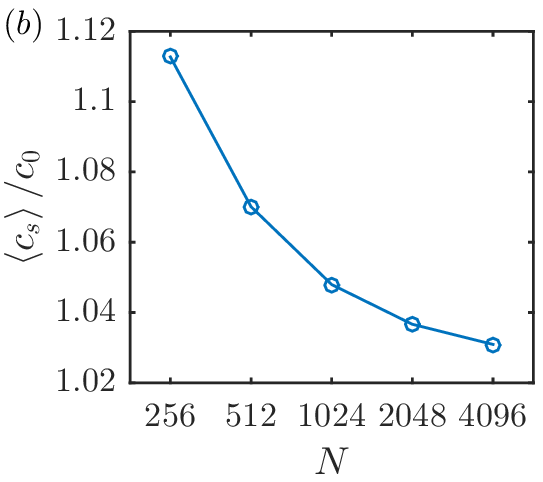} 
        \captionlistentry{}
        \label{fig:refinementStudyCsLowestC0Conv}
    \end{subfigure}
    \caption{Grid convergence study for the lowest value of $c_0$, showing ($a$) $c_s/c_0$ and ($b$) $\mean{c_s}/c_0$. }
    \label{fig:refinementStudyLowestC0}
\end{figure}

Because of the larger differences for the highly skewed distribution, we performed simulations with this $c_0$ using additional refinements giving $N = 512$, $2048$, and $4096$. The resulting distributions of $c_s$ are shown in fig.~\ref{fig:refinementStudyCsLowestC0}, and the values of $\mean{c_s}$ in fig.~\ref{fig:refinementStudyCsLowestC0Conv}. The grid with $N = 1024$ had a relative difference in $\mean{c_s}$ of $1.7\%$ to the finest grid ($N = 4096$). The corresponding difference in the peak of $\theta$ was $2.5\%$. We consider these differences acceptable.

\section{Analytical description of the flow}
\label{sec:analyticalExpressionVelTransGroove}
In the external flow ($y > 0$), Stokes equations \eqref{eq:momentumTransport} are equivalent to the biharmonic equation for the stream function $\psi$,
\begin{equation}
    \nabla^4 \psi = 0, \quad \text{ where } \quad u = \frac{\partial \psi}{\partial y}, \quad v = -\frac{\partial \psi}{\partial x}. 
\end{equation}
Because of symmetry, we can limit the streamwise coordinate to $0 \le x \le p/2$. The streamwise velocity boundary condition is
\begin{equation}
    \frac{\partial \psi}{\partial y} = 0 \quad \text{for} \quad \frac{w}{2} < x \le \frac{p}{2},
    \label{eq:streamFunctionvelocityBC}
\end{equation}
and the shear-stress boundary conditions are 
\begin{align}
    \left.\frac{\partial^2 \psi}{\partial y^2}\right|_{y=0^+} &= \frac{\tau_\mathrm{LIS} + \tau_\MaNumber}{\mu_\infty} &\text{for}& \quad 0 \le x < \frac{w}{2} \quad \text{(eq.~\ref{eq:shearBC})}, \label{eq:streamFunctionShearBC1} \\
    \lim_{y \to \infty}\frac{\partial^2 \psi}{\partial y^2} &= \frac{\tau_\infty}{\mu_\infty} &\text{for}& \quad 0 \le x \le \frac{p}{2}, \label{eq:streamFunctionShearBC2}
\end{align}
In the following, we will use non-dimensional variables $X = 2x/w$, $Y = 2y/w$, $\alpha = (\pi/2)(w/p)$ and $\tilde{\psi} = 8 \mu_\infty \psi/(\tau_\infty w^2)$.

\subsection{An expression for the stream function}
The stream function of Stokes flow over varying no-slip and no-shear stripes was found by \citet{philip72a}. However, it can be generalised to apply to varying no-slip and constant shear-stress stripes \citep{schonecker14}. The relevant form is
\begin{equation}
    \tilde{\psi} = Y^2 + \frac{\tau_\infty - \tau_\mathrm{LIS} - \tau_\MaNumber}{\tau_\infty}\left(-Y^2 + \frac{Y}{\alpha} \mathrm{Im}\left\{\arccos\left( \frac{\cos(\alpha(X + iY)}{\cos(\alpha)}\right)\right\}\right),
\end{equation}
where $\tau_\mathrm{LIS}$ and $\tau_\MaNumber$ are constants, and $\mathrm{Im}\{\}$ designates the imaginary part. With complex coordinates $\Theta = X + iY$ and $\overline{\Theta} = X - iY$, the stream function can be re-written as
\begin{multline}
    \tilde{\psi} = Y^2 + \frac{\tau_\infty - \tau_\mathrm{LIS} - \tau_\MaNumber}{\tau_\infty}\left(-Y^2 + \frac{Y}{\alpha} \mathrm{Im}\left\{\arccos\left( \frac{\cos(\alpha\Theta)}{\cos(\alpha)}\right)\right\}\right) \\ =
       Y^2 + \frac{\tau_\infty - \tau_\mathrm{LIS} - \tau_\MaNumber}{\tau_\infty}\mathrm{Re}\left\{(\overline{\Theta} - \Theta)\frac{1}{2}\left(-\Theta + \frac{1}{\alpha} \arccos\left( \frac{\cos(\alpha\Theta)}{\cos(\alpha)}\right)\right)\right\} \\ = Y^2 + \frac{\tau_\infty - \tau_\mathrm{LIS} - \tau_\MaNumber}{\tau_\infty}\mathrm{Re}\left\{(\overline{\Theta} - \Theta)\tilde{W}(\Theta)\right\},
\end{multline}
where we have defined 
\begin{equation}
    \tilde{W}(\Theta) = \frac{1}{2}\left(-\Theta + \frac{1}{\alpha} \arccos\left( \frac{\cos(\alpha\Theta)}{\cos(\alpha)}\right)\right),
\end{equation}
and $\mathrm{Re}\{\}$ gives the real part.

The wall-normal derivative of the stream function is
\begin{align}
    \frac{\partial \tilde{\psi}}{\partial Y}& = 2Y + \frac{\tau_\infty - \tau_\mathrm{LIS} - \tau_\MaNumber}{\tau_\infty}2\mathrm{Re}\{-i\tilde{W}(\Theta) + Y\tilde{W}'(\Theta)\}, \\ \left.\frac{\partial \tilde{\psi}}{\partial Y}\right|_{y = 0}& = \frac{\tau_\infty - \tau_\mathrm{LIS} - \tau_\MaNumber}{\tau_\infty}2\mathrm{Im}\{\tilde{W}(X)\},
\end{align}
where $\tilde{W}'(\Theta) = \dif \tilde{W}(\Theta)/\dif \Theta$. On the interface ($Y = 0$), $0 \le X < 1$ and $0 \le \alpha X < \alpha$. Also, $0 < \alpha < \pi/2$. In this interval, $\tilde{W}(X)$ is imaginary, and outside ($1 \le X \le \pi/(2\alpha)$), $\tilde{W}(X)$ is real.  Hence, \eqref{eq:streamFunctionvelocityBC} is satisfied. The velocity on the interface is 
\begin{multline}
    u_s = \left.\frac{\partial \psi}{\partial y}\right|_{y = 0} = \frac{\tau_\infty w}{4\mu_\infty}\left.\frac{\partial \tilde{\psi}}{\partial Y}\right|_{Y = 0} \\ = \frac{\tau_\infty w}{4\mu_\infty}\frac{\tau_\infty - \tau_\mathrm{LIS} - \tau_\MaNumber}{\tau_\infty}\frac{1}{\alpha}\mathrm{Im}\left\{\arccos\left( \frac{\cos(\alpha X)}{\cos(\alpha)}\right)\right\} \\ = \frac{\tau_\infty w}{4\mu_\infty}\frac{\tau_\infty - \tau_\mathrm{LIS} - \tau_\MaNumber}{\tau_\infty}\frac{1}{\alpha} \mathrm{arccosh}\left( \frac{\cos(\alpha X)}{\cos(\alpha)}\right).
    \label{eq:interfaceVelocityDistribution}
\end{multline}
The integral \citep{philip72b} 
\begin{equation}
    \int_0^1 \mathrm{arccosh}\left( \frac{\cos(\alpha X)}{\cos(\alpha)}\right) \dif X = -\frac{\pi}{2\alpha}\ln\left(\cos \alpha\right),
\end{equation}
which corresponds to the average of the integrand over the complete interface. The mean slip velocity is therefore
\begin{equation}
    U_s = -\frac{\tau_\infty p}{\mu_\infty 2\pi}\frac{\tau_\infty - \tau_\mathrm{LIS} - \tau_\MaNumber}{\tau_\infty}\ln( \cos \alpha).
    \label{eq:meanSlipVelApp}
\end{equation}
The velocity in the centre of the interface is 
\begin{equation}
    u_s^0 = \frac{\tau_\infty w}{\mu_\infty 4}\frac{\tau_\infty - \tau_\mathrm{LIS} - \tau_\MaNumber}{\tau_\infty}\frac{1}{\alpha}\mathrm{arccosh}\left( \frac{1}{\cos\alpha}\right). 
    \label{eq:middleInterfaceVelApp}
\end{equation}

The second wall-normal derivative of the stream function is
\begin{align}
    \frac{\partial^2 \tilde{\psi}}{\partial Y^2} &= 2 + \frac{\tau_\infty - \tau_\mathrm{LIS} - \tau_\MaNumber}{\tau_\infty}2\mathrm{Re}\left\{2\tilde{W}'(\Theta) + iY W''(\Theta)\right\}, \\
    \left.\frac{\partial^2 \tilde{\psi}}{\partial Y^2}\right|_{Y = 0} &= 2 + \frac{\tau_\infty - \tau_\mathrm{LIS} - \tau_\MaNumber}{\tau_\infty}4\mathrm{Re}\left\{\tilde{W}'(X)\right\},
\end{align}
where $\tilde{W}''(\Theta) = \dif^2 \tilde{W}(\Theta)/\dif \Theta^2$. We can compute 
\begin{equation}
    \tilde{W}'(\Theta) = \frac{1}{2}\left(-1 + \frac{\sin (\alpha\Theta)}{\sqrt{\cos^2 \alpha - \cos^2 (\alpha\Theta)}} \right).
\end{equation}
Hence, for $Y = 0$,
\begin{multline}
    \left.\frac{\partial u}{\partial y}\right|_{y = 0^+} = \left.\frac{\partial^2 \psi}{\partial y^2}\right|_{y = 0} = \frac{\tau_\infty}{2\mu_\infty}\left.\frac{\partial^2 \tilde{\psi}}{\partial Y^2}\right|_{Y = 0} \\ = \frac{\tau_\infty}{\mu_\infty} - \frac{\tau_\infty}{\mu_\infty}\frac{\tau_\infty - \tau_\mathrm{LIS} - \tau_\MaNumber}{\tau_\infty}\mathrm{Re}\left\{1 - \frac{\sin (\alpha X)}{\sqrt{\cos^2 \alpha - \cos^2 (\alpha X)}} \right\}.
\end{multline}
On the interface, $\sqrt{\cos^2 \alpha - \cos^2 (\alpha X)}$ is imaginary, and thereby the boundary condition \eqref{eq:streamFunctionShearBC1} is satisfied. However, the average shear stress at the surface ($0 \le x \le p/2$) is $\tau_\infty$, found by using $\tilde{W}(1) = -1/2$ and $\tilde{W}(p/w) = 0$. 

The first streamwise derivative of $\tilde{\psi}$ is
\begin{equation}
    \frac{\partial \tilde{\psi}}{\partial X} = \frac{\tau_\infty - \tau_\mathrm{LIS} - \tau_\MaNumber}{\tau_\infty}2Y\mathrm{Im}\left\{\tilde{W}'(\Theta)\right\}.
\end{equation}
This expression is zero for $Y = 0$, showing that the wall-normal velocity is zero at the surface. 

\subsection{Modelling the infusing-liquid shear stress}
The interfacial shear stresses of the infusing liquid can be assumed to relate to the interface velocity by a local slip length $\zeta(x)$ \citep{schonecker14}, 
\begin{equation}
    u_s = \frac{\zeta(x)}{\mu_i}\tau_\mathrm{LIS},
\end{equation}
where 
\begin{equation}
    \zeta(x) = \frac{w\mu_i}{4\mu_\infty}C_t\frac{1}{\alpha}\mathrm{arccosh}\left( \frac{\cos(\alpha X)}{\cos(\alpha)}\right) \quad \text{and} \quad C_t = \frac{8\alpha D_t\mu_\infty/\mu_i}{\ln\left(\dfrac{1 + \sin(\alpha)}{1 - \sin(\alpha)}\right)}.
    \label{eq:Ct}
\end{equation}
Here, $D_t = d_t/w$ is the maximum value of $\zeta(x)/w$. 
Together with eq.~\eqref{eq:interfaceVelocityDistribution},
\begin{equation}
    \tau_\mathrm{LIS} = \frac{\tau_\infty}{1 + C_t}\left(1 - \frac{\tau_\MaNumber}{\tau_\infty}\right).
    \label{eq:LIS-MaDependence}
\end{equation}
Hence, $\tau_\infty/(1 + C_t)$ is the interfacial shear stress of the infusing liquid in the absence of surfactants ($\tau_\MaNumber = 0$).

The non-dimensional slip length $D_t$ can be modelled by
\begin{equation}
    D_t = f(a)\beta~\mathrm{erf}\left(\dfrac{g(a)\sqrt{\pi}}{8f(a)\beta}A\right),
    \label{eq:DtModel}
\end{equation}
where
\begin{equation}
    f(a) = -\frac{\ln\left(\dfrac{1 + \sin\left(\dfrac{\pi a}{2}\right)}{1 - \sin\left(\dfrac{\pi a}{2}\right)}\right)}{2a\ln 2 \left(1 + \dfrac{2\ln\left(\cos\left(\dfrac{\pi a}{2}\right)\right)}{2a~\mathrm{arctanh}(a) + \ln(1 - a^2)}\right)}, \quad g(a) = \frac{4}{\pi} - \frac{4 - \pi}{\pi}a,
\end{equation}
$\beta = 0.505/(2\pi)$, $A = k/w$, $a = w/p$, and $\mathrm{erf}(x)$ is the error function.

\subsection{Effective slip length}
The effective slip length $b$ is (using eqs.~\ref{eq:meanSlipVelApp} and \ref{eq:LIS-MaDependence})
\begin{multline}
    b = \frac{\mu_\infty}{\tau_\infty}U_s = -\frac{\mu_\infty}{\tau_\infty}\frac{\tau_\infty p}{\mu_\infty 2\pi}\frac{\tau_\infty - \tau_\mathrm{LIS} - \tau_\MaNumber}{\tau_\infty}\ln( \cos \alpha) \\ = -\frac{p}{2\pi}\left(1 - \frac{1}{1 + C_t}\left(1 - \frac{\tau_\MaNumber}{\tau_\infty}\right) - \frac{\tau_\MaNumber}{\tau_\infty}\right)\ln( \cos \alpha) \\ = -\frac{p}{2\pi}\ln( \cos \alpha)\frac{C_t}{1 + C_t}\left(1 - \frac{\tau_\MaNumber}{\tau_\infty}\right) = 
    b_\mathrm{SHS}\beta_\mathrm{LIS} \left(1 - \frac{\tau_\MaNumber}{\tau_\infty}\right),
    \label{eq:derivedSlipRelation}
\end{multline}
where $b_\mathrm{SHS} = -p\ln(\cos \alpha)/(2\pi)$ is the slip length for $\mu_i/\mu_\infty \to 0$ and $\tau_\MaNumber = 0$, and $\beta_\mathrm{LIS} = C_t/(1 + C_t)$ expresses the effects of the viscous infusing liquid. For SHS with $\mu_i/\mu_\infty \to 0$, $\beta_\mathrm{LIS} \to 1$. The velocity at the centre of the interface (eq.~\ref{eq:middleInterfaceVelApp}) becomes
\begin{equation}
    u_{s}^0 = \frac{\tau_\infty w}{4\mu_\infty\alpha}\mathrm{arccosh}\left( \frac{1}{\cos\alpha}\right)\frac{C_t}{1 + C_t}\left(1 - \frac{\tau_\MaNumber}{\tau_\infty}\right) = u_{s,\mathrm{SHS}}^0\beta_\mathrm{LIS}\left(1 - \frac{\tau_\MaNumber}{\tau_\infty}\right).
    \label{eq:derivedMiddleInterfaceVel}
\end{equation}

\section{Derivation of surfactant boundary layer thickness}
\label{sec:derivationOfDelta}
In this section, we estimate $\delta/w$. The limits of low and high $\PeNumber$ are considered separately, and the final expression is formed as a combination of these two limits.

\subsection{Low P\'eclet numbers}
For low $\PeNumber$, advection is assumed to be negligible, corresponding to (cf.~\ref{eq:bulkTransport})
\begin{equation}
    \frac{\partial^2 c}{\partial x^2} + \frac{\partial^2 c}{\partial y^2} = 0.
    \label{eq:bulkTransportLowPe}
\end{equation}
We integrate this equation one time in the streamwise and one time in the wall-normal direction. 

For the wall-normal diffusion (second term) in the wall-normal direction,
\begin{equation}
    \int_0^\infty \frac{\partial^2 c}{\partial y^2} \dif y = \lim_{y \to \infty}\frac{\partial c}{\partial y} - \left. \frac{\partial c}{\partial y}\right|_{y=0} \approx 0 - \frac{\Delta c_s}{\delta}f_y(x),
    \label{eq:wallNormalDiffWallNormalInt}
\end{equation}
where $f_y(x) = -2x/w$ for $-w/2 < x < w/2$ and $0$ otherwise, using the assumed spatial dependency of eq.~\eqref{eq:DeltacEstimation} and eq.~\eqref{eq:cGradScaling} for the value at $x = -w/2$. Integrating the result in the streamwise direction from $-p/2$ to $0$,
\begin{equation}
    \int_{-p/2}^0 - \frac{\Delta c_s}{\delta}f_y(x) \dif x = \int_{-w/2}^0 - \frac{\Delta c_s}{\delta}\frac{-2x}{w} \dif x =  -\frac{\Delta c_s}{\delta}\frac{w}{4}.
    \label{eq:lowPeDeltaWallnormalDiff}
\end{equation}

For the streamwise diffusion, we need to assume a wall-normal dependency of $c$. We assume that 
\begin{equation}
\begin{array}{lll}
c &= c_0 - \Delta c_s f_x(x)\dfrac{\delta - y}{\delta}  \quad &\text{for $0 < y < \delta$}, \\
c &= 0 \quad &\text{otherwise},
\end{array}
\label{eq:streamwiseDiffSpatialDependency}
\end{equation}
and $f_x(x)$ describes the streamwise dependency with $f_x(-w/2) = 1$ and $f_x(0) = f_x(-p/2) = 0$ (similarly to $f_y(x)$). The streamwise dependency is consistent with eq.~\eqref{eq:DeltacEstimation}. The wall-normal dependency is consistent with eq.~\eqref{eq:cGradScaling} but not correct outside the interface ($x > w/2$ and $x < -w/2$) since there the wall-normal derivative at the wall needs to be zero. However, we think that it is a good enough approximation. Thus,
\begin{equation}
    \int_0^\infty \frac{\partial^2 c}{\partial x^2} \dif y = \int_0^\delta \frac{\partial^2 c}{\partial x^2} \dif y = - \frac{1}{2}\Delta c_s f_x''(x)\delta,
\end{equation}
where $f_x''(x) =  \dif^2 f_x(x)/\dif x^2$. It can be noted that the alternative wall-normal distribution $\exp(-y/\delta)$ would give a coefficient of $1$ instead of $1/2$. Integrating the result in the streamwise direction,
\begin{multline}
    - \frac{1}{2}\Delta c_s \delta \int_{-p/2}^0 \frac{\dif^2 f_x}{\dif x^2} \dif x = - \frac{1}{2}\Delta c_s \delta\left(f_x'(0) - f_x'(-p/2)\right) \approx \\ - \frac{1}{2}\Delta c_s \delta\left(\frac{f_x(0) - f_x(-w/2)}{w/2} - \frac{f_x(-w/2) - f_x(-p/2)}{(p - w)/2}\right) = \Delta c_s \delta\left(\frac{1}{w} + \frac{1}{p - w}\right),
    \label{eq:lowPeDeltaStreamwiseDiff}
\end{multline}
where $f_x'(x) =  \dif f_x(x)/\dif x$.

Using eqs.~\eqref{eq:lowPeDeltaWallnormalDiff} and \eqref{eq:lowPeDeltaStreamwiseDiff} in eq.~\eqref{eq:bulkTransportLowPe},
\begin{equation}
    \frac{\delta}{w} = \frac{1}{2}\sqrt{1 - \frac{w}{p}}.
    \label{eq:derivedDeltaLowPe}
\end{equation}

Contrary to the expression by \citet{landel20}, eq.~\eqref{eq:derivedDeltaLowPe} accounts for diffusion between boundary layers of adjacent interfaces. This diffusion is relevant when the solid fraction is low ($w \gtrsim p-w$).

\subsection{High P\'eclet numbers}
For high $\PeNumber$, we assume that streamwise advection balances wall-normal diffusion,
\begin{equation}
    u\frac{\partial c}{\partial x} = D\frac{\partial^2 c}{\partial y^2}.
    \label{eq:bulkTransportHighPe}
\end{equation}
We also integrate this equation in the wall-normal and streamwise directions. The integration of the wall-normal diffusion is the same as above. For the advection, we retain the wall-normal dependency of eq.~\eqref{eq:streamwiseDiffSpatialDependency} but replace $f_x(x)$ with another function, $g_x(x)$. Since we neglect streamwise diffusion, we assume that the concentration at the stagnation point of the upstream interface is advected downstream, $g_x(-p/2) = -1$. However, we consider $g_x(0) = f_x(0) = 0$. Similar to \citet{landel20}, we assume the L\'ev\^eque regime: $u = y\tau_\infty/\mu_\infty$. Thus,
\begin{equation}
    \int_0^\infty u\frac{\partial c}{\partial x} \dif y = -\frac{\tau_\infty}{\mu_\infty}\Delta c_s g_x'(x) \int_0^\delta y\frac{\delta - y}{\delta} \dif y  = -\frac{1}{6}\frac{\tau_\infty}{\mu_\infty}\Delta c_s g_x'(x)\delta^2.
\end{equation}
Performing the streamwise integration, 
\begin{multline}
    - \frac{1}{6}\frac{\tau_\infty}{\mu_\infty}\Delta c_s \delta^2 \int_{-p/2}^0 g_x'(x) \dif x = - \frac{1}{6}\frac{\tau_\infty}{\mu_\infty}\Delta c_s \delta^2 (g_x(0) - g_x(-p/2)) \\ = -\frac{1}{6}\frac{\tau_\infty}{\mu_\infty}\Delta c_s \delta^2 = -\frac{1}{6}\PeNumber\Delta c_s D\frac{\delta^2}{w^2}.
    \label{eq:highPeDeltaStreamwiseAdv}
\end{multline}

Using eqs.~\eqref{eq:lowPeDeltaWallnormalDiff} and \eqref{eq:highPeDeltaStreamwiseAdv} in eq.~\eqref{eq:bulkTransportHighPe},
\begin{equation}
    \frac{\delta}{w} = \left(\frac{3}{2}\frac{1}{\PeNumber} \right)^{1/3}.
    \label{eq:derivedDeltaHighPe}
\end{equation}

\subsection{Combined expression}
The two expressions \eqref{eq:derivedDeltaLowPe} and \eqref{eq:derivedDeltaHighPe} can be combined into a single expression with the correct asymptotic behaviours,
\begin{equation}
    \frac{\delta}{w} = \frac{\frac{1}{2}\sqrt{1 - \frac{w}{p}}}{\left(1 + \frac{2}{3}\PeNumber\left(\frac{1}{2}\sqrt{1 - \frac{w}{p}}\right)^3\right)^{1/3}}.
    \label{eq:derivedDelta}
\end{equation}

\begin{figure}
    \centering
    \begin{subfigure}{0.4\textwidth}
        \centering
        \includegraphics[height=4.5cm]{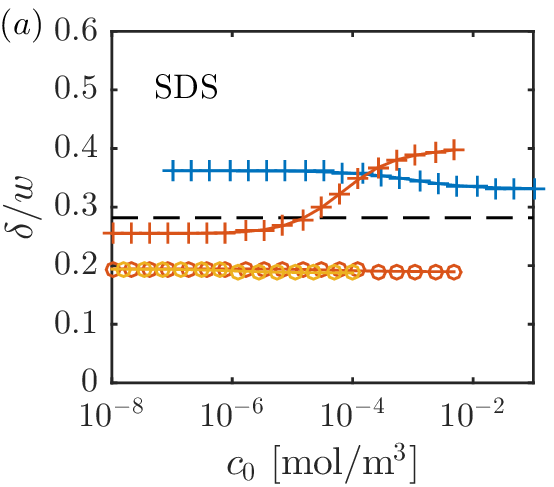} 
        \captionlistentry{}
        \label{fig:deltaComparisonSDS}
    \end{subfigure}
    \begin{subfigure}{0.4\textwidth}
        \centering
        \includegraphics[height=4.5cm]{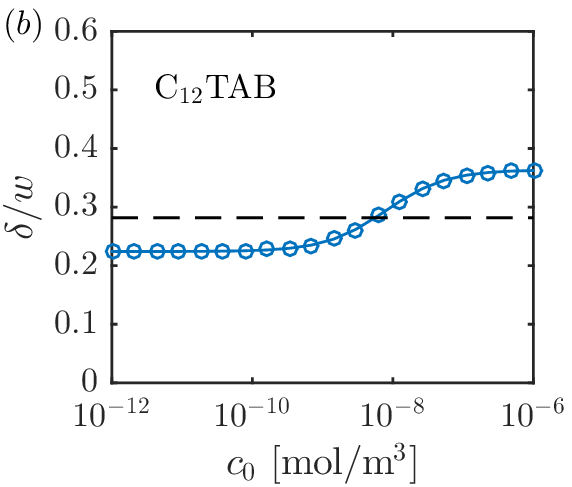} 
        \captionlistentry{}
        \label{fig:deltaComparisonC12TAB}
    \end{subfigure}
    \begin{subfigure}{0.4\textwidth}
        \centering
        \includegraphics[height=4.5cm]{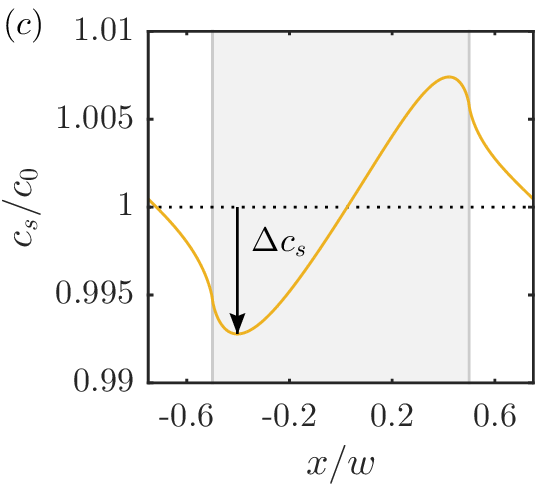} 
        \captionlistentry{}
        \label{fig:frumkinSDSDeltaCS}
    \end{subfigure}
    \begin{subfigure}{0.4\textwidth}
        \centering
        \includegraphics[height=4.5cm]{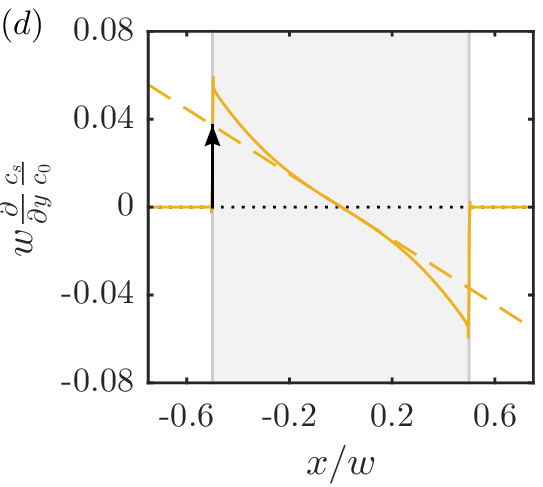} 
        \captionlistentry{}
        \label{fig:frumkinfrumkinSDSdcdy}
    \end{subfigure}
    \caption{Comparison of $\delta/w$ computed from eq.~\eqref{eq:derivedDelta} (\longbroken) with estimations from simulations using ($a$) SDS (colours and symbols as in fig.~\ref{fig:frumkinSDSComparison}) and ($b$) C$_{12}$TAB. The simulation estimates are based on the left relationship of eq.~\eqref{eq:cGradScaling}, with $\Delta c_s$ and $\left.\partial c/\partial y\right|_{y = 0, x = -w/2}$ as the arrows in ($c$) and ($d$), respectively. The specific distributions in these two figures are from the lowest simulated SDS concentration using the Frumkin model with alkane-surfactant interaction, $c_0 = 10^{-10}$ mol/m$^3$. The grey boxes in ($c$) and ($d$) indicate the interface, and the dotted lines correspond to $c_s/c_0 = 1$ and $\partial c_s/\partial y = 0$, respectively. The dashed line in ($d$) is a linear extrapolation from the centre of the interface. }
    \label{fig:deltaEstimation}
\end{figure}

In fig.~\ref{fig:deltaEstimation}, eq.~\eqref{eq:derivedDelta} is compared to values of $\delta/w$ from simulations of secs.~\ref{sec:regularFrumkin} and \ref{sec:advancedFrumkin}, estimated using the left relation of \eqref{eq:cGradScaling}. The value of $\Delta c_s$ was computed using the minimum value of $c_s$ (fig.~\ref{fig:frumkinSDSDeltaCS}). The concentration wall-normal derivative showed significant non-linear behaviour close to the interface edges. Therefore, a reasonable value was determined by extrapolating a line to $x = -w/2$ with the same slope as around $x = 0$ (fig.~\ref{fig:frumkinfrumkinSDSdcdy}). 
There is a satisfactory agreement between the prediction and the simulation results.

For the simulations with a higher $\PeNumber$ (sec.~\ref{sec:stagnantCap}), eq.~\eqref{eq:derivedDelta} gives $\delta/w = 0.068$. The assumption of linear interfacial surfactant distributions is valid for the highest simulated concentration, giving an estimate of $\delta/w = 0.034$. Even if the relative difference between prediction and estimation is larger for the higher $\PeNumber$, it does not influence the conclusions in any significant way.

\section{Reinterpretation of data from Landel \textit{et al.} (2020)}
\label{sec:landel2020}
\begin{figure}
    \centering
    \begin{subfigure}{0.49\textwidth}
        \centering
        \includegraphics[height=5cm]{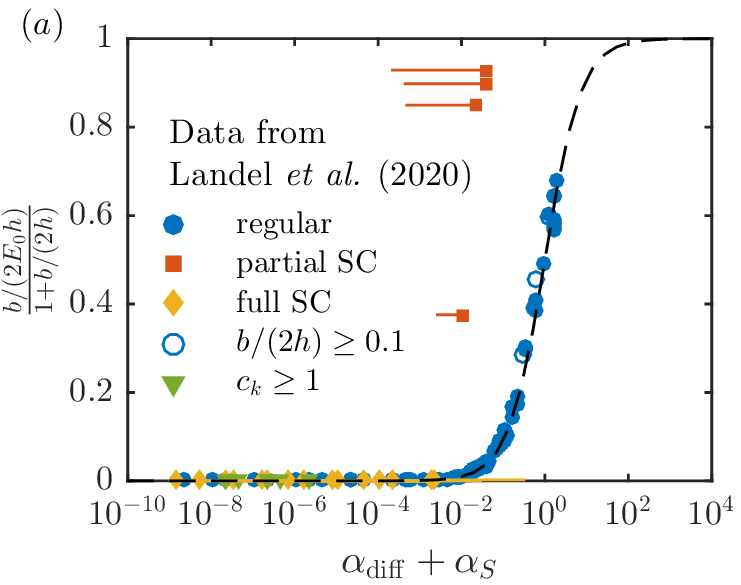} 
        \captionlistentry{}
        \label{fig:landel2020AlphaSlipPlot}
    \end{subfigure}
    \begin{subfigure}{0.49\textwidth}
        \centering
        \includegraphics[height=5cm]{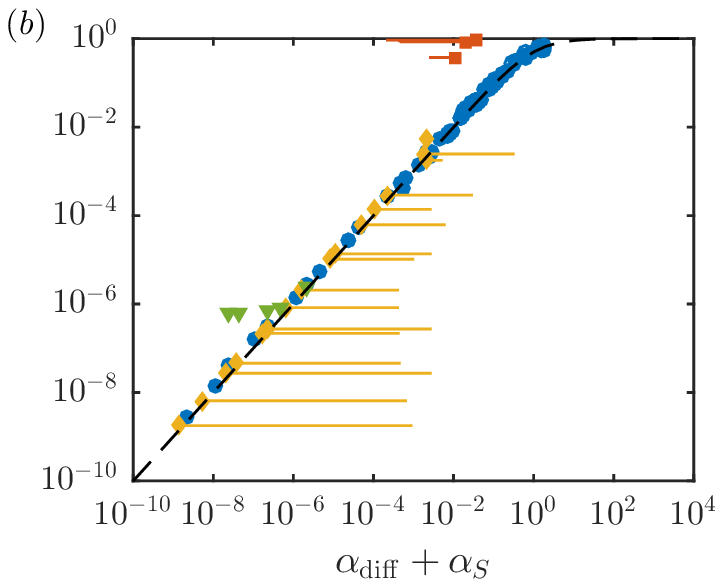} 
        \captionlistentry{}
        \label{landel2020AlphaSlipPlotLog}
    \end{subfigure}
    \begin{subfigure}{0.49\textwidth}
        \centering
        \includegraphics[height=5cm]{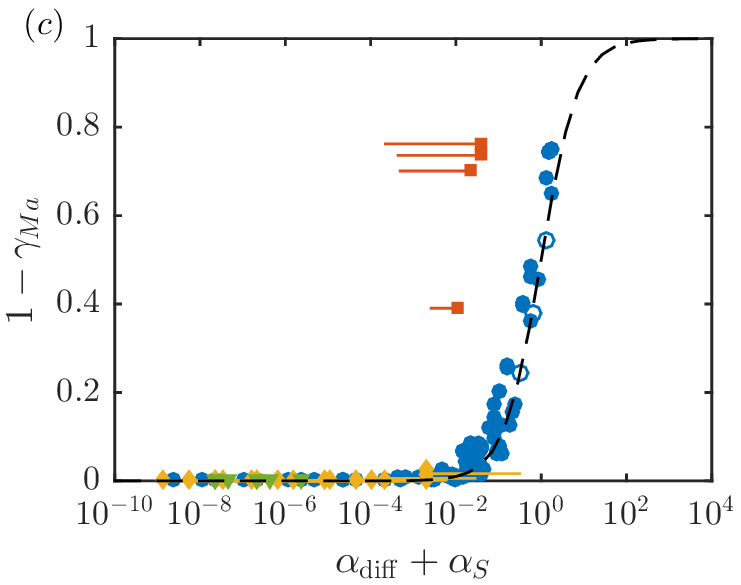} 
        \captionlistentry{}
        \label{landel2020AlphaGammaPlot}
    \end{subfigure}
    \caption{The analytical model (\longbroken) compared to data from \citet{landel20}, showing $b$ (by eq.~\ref{eq:landelSlipLength}) with ($a$) linear and ($b$) logarithmic vertical axis, and ($c$) $1 - \gamma_\MaNumber$. Points in the partial and full SC regimes and points with $b/(2h) \ge 0.1$ or $c_k \ge 1$ have been identified. Other points have been denoted as regular. For results in the partial or full SC regime, lines go from the points to the values of $\alpha_\mathrm{diff} + \alpha_S$ given by eq.~\eqref{eq:linearLimit}.}
    \label{fig:landel2020}
\end{figure}

Using the analytical model derived in this article, we reinterpreted the data from simulations of SHS ($\beta_\mathrm{LIS} = 1$) with surfactants in two-dimensional Poiseuille flow from \citet{landel20}. These results are shown in fig.~\ref{fig:landel2020}, showing both $b$ and $\gamma_\MaNumber$, where the latter is the averaged shear stress on the interface normalised by the average shear stress of the channel, corresponding to $\tau_\MaNumber/\tau_\infty$. Therefore, $\gamma_\MaNumber$ can be compared to the analytical model of eq.~\eqref{eq:slipLengthEstimation} (corresponding to eq.~4.29 of \citealt{landel20}),
\begin{equation}
    1 - \gamma_\MaNumber = 1 - \frac{1}{\alpha_\mathrm{diff} + \alpha_S + 1},
\end{equation}
with $\alpha_\mathrm{diff}$ and $\alpha_S$ from eq.~\eqref{eq:alpha}. However, because of the different geometry, $b$ is not proportional to $1 - \gamma_\MaNumber$, as eq.~\eqref{eq:slipRelation} states for the geometry of the current work. Instead,
\begin{equation}
    b = \frac{2(1 - \gamma_\MaNumber)E_0}{1 - (1 - \gamma_\MaNumber)E_0}h \quad \implies \quad \frac{b/(2hE_0)}{1 + b/(2h)} = 1 - \gamma_\MaNumber,
    \label{eq:landelSlipLength}
\end{equation}
for a channel half-height $h$ and the geometry-dependent constant $E_0$ (eq.~4.34 of \citealt{landel20}). For $b \ll 2h$, eqs.~\eqref{eq:slipRelation} and \eqref{eq:landelSlipLength} give $b_\mathrm{SHS} \approx 2hE_0$. We plot the full expression \eqref{eq:landelSlipLength}, marking the three points where non-linear geometrical effects are present ($b/(2h) \ge 0.1$).

The total number of points is 137. In the figure, we have marked data points in the partial (4 points) and full SC (16 points) regimes as identified by \citet{landel20} and added a line going from the simulation results to $\alpha_\mathrm{diff} + \alpha_S$ of eq.~\eqref{eq:linearLimit}. For interfaces in the partial SC regime, the simulation point is to the right of this value, whereas for interfaces in the full SC regime, the simulation points are to the left. This behaviour agrees with eq.~\eqref{eq:linearLimit}, as no linear distributions of $\theta$ can exist above this $\alpha_\mathrm{diff} + \alpha_S$, and the SC becomes partial. In addition to the SC points, we have identified five cases with $c_k \ge 1$. None of these simulations resulted in $b/(2h) \ge 0.1$.

\citet{landel20} did not use eq.~\eqref{eq:requirementForExistenceOfPSC} to distinguish between uniform and SC regimes. All SC interfaces identified in fig.~\ref{fig:landel2020} did fulfil \eqref{eq:requirementForExistenceOfPSC}. However, some points classified as uniformly retarded fulfilled \eqref{eq:requirementForExistenceOfPSC} as well. Out of these, 6 points should be in the partial SC regime according to eq.~\eqref{eq:linearLimit}. The non-linearity of their interfacial surfactant distributions should be visible upon inspection. However, with their $C_\mathrm{SC}$ relatively close to $1$ (minimum $C_\mathrm{SC} = 0.15$, cf.~$0.059$ in \ref{eq:currentParameters}), only a minor deviation from the predicted slip length is allowed; a smaller $C_\mathrm{SC}$ gives a lower $\alpha_\mathrm{diff} + \alpha_S$ limit where the analytically model becomes valid, allowing $b$ to deviate further (eq.~\ref{eq:linearLimit} and fig.~\ref{fig:033SDSAlphaMap}). Therefore, these SC presumably had a negligible influence on the results in fig.~\ref{fig:landel2020}.

The analytical model seems to fit $b$ better than $\gamma_\MaNumber$. This observation appears to be consistent with what is reported by \citet{landel20}; $\gamma_\MaNumber$ is probably affected by numerical errors to a higher degree than $b$. It can also be noted that $u_{s,\mathrm{SHS}}^0$ (eq.~\ref{eq:middleInterfaceVel}) corresponds to $2\mathcal{F}_0$ of \citet{landel20}. 
It is exactly their leading-order expression for narrow grooves (eq.~C24). 

\bibliographystyle{jfm} \bibliography{references}
\end{document}